\documentclass{emulateapj}

\def\bm#1{\mbox{\boldmath $#1$}} 
\newcommand{\comments}[1]{}
\newcommand{\ba}{\begin{eqnarray}}
\newcommand{\ea}{\end{eqnarray}}
\newcommand{\be}{\begin{equation}}
\newcommand{\ee}{\end{equation}}
\newcommand{\lan}{\langle}
\newcommand{\ran}{\rangle}
\newcommand{\grad}{\nabla}

\begin{document}

\title{Buoyancy Instabilities in Galaxy Clusters: Convection Due to Adiabatic Cosmic Rays and Anisotropic Thermal Conduction}

\author{Prateek Sharma\altaffilmark{1}}
\affil{Theoretical Astrophysics Center and Astronomy Department, University of California,
    Berkeley, CA 94720}
\email{psharma@astro.berkeley.edu}

\author{Benjamin D. G. Chandran}
\affil{Space Science Center and Department of Physics, University of New Hampshire, Durham, NH 03824}
\email{benjamin.chandran@unh.edu}

\author{Eliot Quataert, Ian J. Parrish\altaffilmark{1}}
\affil{Theoretical Astrophysics Center and Astronomy Department, University of California,
    Berkeley, CA 94720}
\email{eliot@astro.berkeley.edu, iparrish@astro.berkeley.edu}
\altaffiltext{1}{Chandra Fellow}

\begin{abstract}
  Using a linear stability analysis and two and three-dimensional
  nonlinear simulations, we study the physics of buoyancy
  instabilities in a combined thermal and relativistic (cosmic ray)
  plasma, motivated by the application to clusters of galaxies.  We
  argue that cosmic ray diffusion is likely to be slow compared to the
  buoyancy time on large length scales, so that cosmic rays are
  effectively adiabatic.  If the cosmic ray pressure $p_{cr}$ is
  $\gtrsim 25 \%$ of the thermal pressure, and the cosmic ray entropy
  ($p_{\rm cr}/\rho^{4/3}$; $\rho$ is the thermal plasma density)
  decreases outwards, cosmic rays drive an adiabatic convective
  instability analogous to Schwarzschild convection in stars.  Global
  simulations of galaxy cluster cores show that this instability
  saturates by reducing the cosmic ray entropy gradient and driving
  efficient convection and turbulent mixing.  At larger radii in
  cluster cores, the thermal plasma is unstable to the heat
  flux-driven buoyancy instability (HBI), a convective instability
  generated by anisotropic thermal conduction and a background
  conductive heat flux.  The HBI saturates by rearranging the magnetic
  field lines to become largely perpendicular to the local
  gravitational field; the resulting turbulence also primarily mixes
  plasma in the perpendicular plane. Cosmic-ray driven convection and
  the HBI may contribute to redistributing metals produced by Type 1a
  supernovae in clusters.  Our calculations demonstrate that adiabatic
  simulations of galaxy clusters can artificially suppress the mixing
  of thermal and relativistic plasma; anisotropic thermal conduction
  allows more efficient mixing, which may contribute to cosmic rays
  being distributed throughout the cluster volume.

\end{abstract}


\section{Introduction}

\label{intro}

Microscopic transport of heat and momentum in dilute plasmas, like
those in clusters of galaxies, is primarily along magnetic field lines
\citep[][]{bra65}. This anisotropic transport dramatically affects the
convective stability of the plasma; convective stability is no longer
determined by the entropy gradient \citep[][]{sch58}. Instead, a
plasma is unstable to buoyant motions irrespective of the background
entropy and temperature gradients \citep[][]{bal00,qua08}. Cosmic rays
diffusing along magnetic field lines also affect the convective
stability of the plasma \citep[][]{cha06,den08}. Nonlinear simulations
show that these instabilities driven by anisotropic thermal and cosmic
ray transport can change the magnetic field configuration, and the
background temperature and density profiles in the plasma, but they do
not drive efficient convection
\citep[e.g.,][]{par07,par08a,par08b,sha08}. The instabilities saturate
largely by rearranging the magnetic field configuration, thereby
slowing down the instability and reaching a state of marginal
stability to linear perturbations. By contrast, hydrodynamic
convection in stellar interiors redistributes energy efficiently to
make the plasma nearly adiabatic and thus marginally stable to
convection.

One of the key astrophysical motivations for studying the transport
properties of dilute plasmas in the presence of cosmic rays is to
understand the dynamical and thermal structure of clusters of
galaxies. The radiative cooling time ($\lesssim 1$ Gyr) is much less
than the Hubble time ($t_H \approx 13.7$ Gyr) in cluster cores. Thus,
it was expected that the intracluster medium (ICM) would cool rapidly,
resulting in large rates ($\gtrsim 100 M_\odot$ yr$^{-1}$) of mass
cooling to form cold gas and stars \citep[e.g.,][]{fab94}. However,
X-ray observations have failed to detect copious emission from the
expected cold plasma component in cluster cores
\citep[e.g.,][]{pet03}. The lack of cooling flows implies that cooling
is balanced by some source of heating, e.g., heating by thermal
conduction from large radii \citep[][]{ber86,zak03}, heating by jets
and bubbles blown by a central AGN \citep[][]{bin95,cio01}, or heating
by cosmic rays \citep[e.g.,][]{ros89,loe91,cha07}. Although thermal
conduction may operate at large radii, it appears that the plasma at
small radii must be heated by a feedback process which efficiently
self-regulates. This is required to avoid the fine tuning of thermal
conductivity required in models that include only conduction
\citep[e.g.,][]{guo08,con08}.

Cosmic rays from a central AGN have been invoked to prevent
catastrophic cooling of the plasma in cluster cores, either directly
via Alfv\'en waves driven by cosmic rays heating the plasma
\citep[e.g.,][]{loe91,guo08}, or indirectly via convection driven by
cosmic rays, the dissipation of which heats the plasma
\citep[][]{cha07,reb06}. Although there is ample evidence for the
presence of cosmic rays in radio emitting bubbles in clusters
\citep[e.g.,][]{bir04}, it is unclear how/whether cosmic rays can be
spread throughout the cluster volume at a sufficient level for these
heating mechanisms to work. Simple hydrodynamic jets do not couple
their energy to most of the ICM and instead simply drill through it,
without heating and without transporting cosmic rays (and metals)
throughout the ICM \citep[][]{ver06}.

In this paper we show that cosmic rays, which are likely to be
centrally concentrated in clusters, can drive efficient convection and
mixing if the cosmic ray pressure is not negligible compared to the
plasma pressure ($p_{\rm cr}/p \gtrsim 0.25$); this is likely the case
in and around radio bubbles.  
We argue that
on the large scales that likely dominate the turbulent dynamics in the
ICM, cosmic rays are effectively adiabatic rather than diffusive
(i.e., the cosmic ray diffusion time is longer than the buoyancy
time).  As a result, the cosmic rays can drive a Schwarzschild-like
adiabatic convective instability.  We present a linear analysis
demonstrating that, while magnetic reorientation can shut off
diffusive (``isobaric'') cosmic ray instabilities, it cannot shut off
the adiabatic buoyancy instability driven by a negative cosmic ray
entropy gradient. We then present two-fluid (plasma and cosmic rays)
numerical simulations with thermal conduction and cosmic ray diffusion
along magnetic field lines.

We do {\em not} include plasma cooling in this paper, nor do we
include the heating of the thermal plasma that arises from the
excitation of short wavelength Alfv\'en waves by streaming cosmic rays.  
These are both significant omissions and preclude our results
from being an accurate representation of the plasma in cluster cores.  However,
the main focus of this paper is not to solve the cooling flow problem
per se, but rather to isolate and understand the transport and
turbulence properties of cluster plasmas with realistic physics (e.g.,
anisotropic conduction, convection, and cosmic rays).
By neglecting cooling, our calculations implicitly assume that some
unspecified source of heating is preventing the rapid cooling of the
ICM.  A study of cluster cores with cooling and anisotropic conduction
will be presented in a separate paper.

The remainder of this paper is organized as follows. In \S2 we present
the basic equations used in our analysis and derive the dispersion
relation for linear buoyancy waves in the presence of thermal plasma
and cosmic rays.  In \S3 we describe our numerical simulations and
show that a steep cosmic-ray entropy gradient drives convective
motions at small radii in cluster cores, and that anisotropic thermal
conduction drives convection at intermediate radii that rearranges the
magnetic field structure in clusters.  Readers interested in just the
numerical results can skip the linear stability calculation in \S 2.
In \S4 we summarize and discuss the implications of our work.

\section{Basic Equations}
To describe cosmic rays and thermal plasma in the ICM, we use the
two-fluid model of \citet{dru81}, modified to include anisotropic
transport, gravitational acceleration $\bm{g} = - g\bm{\hat {r}}$ 
($g=d\Phi/dr$, where $\Phi$ is the gravitational potential), and a cosmic ray source term to
build up cosmic ray pressure. The
equations of this model are 
\begin{equation}
\frac{d \rho}{d t} = - \rho \grad \cdot \bm{v},
\label{eq:cont}
\end{equation}
\begin{equation}
\rho \frac{d\bm{v}}{dt} = \frac{(\grad \times \bm{B}) \times \bm{B}}{4\pi}
- \grad (p + p_{\rm cr}) - \rho g \bm{\hat{r}},
\label{eq:momentum}
\end{equation}
\begin{equation}
\frac{\partial \bm{B}}{\partial t} = \grad \times (\bm{v} \times \bm{B}),
\label{eq:ind}
\end{equation}
\begin{equation}
\frac{dp}{dt} - \frac{\gamma p}{\rho} \,\frac{d\rho}{dt} 
= - (\gamma - 1)\grad \cdot \bm{Q},
\label{eq:energy}
\end{equation}
and
\begin{equation}
\frac{dp_{\rm cr}}{dt} - \frac{\gamma_{\rm cr} p_{\rm cr}}{\rho} 
\,\frac{d\rho}{dt} 
= - \grad \cdot \bm{\Gamma} + (\gamma_{\rm cr}-1) Q_c
\label{eq:crenergy}
\end{equation}
where $d/dt = \partial/\partial t +
\bm{v} \cdot \grad$ is the Lagrangian time derivative,
\begin{equation}
\bm{Q} = - \kappa_\parallel \bm{ \hat{b}(\hat{b} \cdot \grad} T)
\label{eq:defQ}
\end{equation}
is the heat flux, $T$ is the plasma temperature,
\begin{equation}
\bm{\Gamma} = - D_\parallel \bm{\hat{b}(\hat{b}} \cdot \grad
  p_{\rm cr})
\label{eq:defGamma}
\end{equation}
is the diffusive flux of cosmic-ray energy (multiplied by
$[\gamma_{\rm cr}-1]$), $Q_c$ is the cosmic ray energy source term,
$\rho$ is the mass density, $\bm{v}$ is the common bulk-flow velocity
of the thermal plasma and cosmic rays, $\bm{B}$ is the magnetic field,
$\bm{\hat{b}} = \bm{B}/B$, $p$ and $p_{\rm cr}$ are the thermal-plasma
and cosmic-ray pressures, $\kappa_\parallel$ is the parallel thermal
conductivity, $D_\parallel$ is the diffusion coefficient for
cosmic-ray transport along the magnetic field, and $\gamma$=5/3 and
$\gamma_{\rm cr}$=4/3 are the adiabatic indices of the thermal plasma
and cosmic rays, respectively.

As mentioned in \S \ref{intro}, we do not include radiative cooling in
the energy equation (eq. [\ref{eq:energy}]).  In addition, we do not
include the effects of cosmic ray streaming relative to the thermal
plasma: in particular we neglect Alfv\'en wave heating of the thermal
plasma and Alfv\'en wave streaming in the cosmic ray energy equation
\citep[e.g.,][]{loe91}.  This physics will be included in the future
together with plasma cooling.  For the present paper we focus on the
physics of convective instabilities in clusters using a simplified but
physically reasonable model.

\subsection{Linear Stability Analysis}

We take all quantities to be the sum of an equilibrium value plus a
small-amplitude fluctuation: $\bm{B} = \bm{B}_0 + \bm{B}_1$, etc. We
take the equilibrium velocity~$\bm{v}_0$ to vanish, set $\bm{B}_0 =
B_{0x} \bm{\hat{x}} + B_{0z}\bm{\hat{z}}$ ($\bm{\hat{z}} =
\bm{\hat{r}}$ for the cluster, and $\bm{\hat{x}}$ is chosen such that
local magnetic field lies in the $\bm{\hat x}-\bm{\hat{z}}$ plane),
and take $T_0$, $p_{\rm cr0}$, and $\rho_0$ to be functions of
$z$~alone. We employ a local analysis in which all fluctuating
quantities vary as $e^{i\bm{k} \cdot \bm{x} - i \omega t}$ with $kH
\gg 1$, where $H$ is the scale on which the equilibrium quantities
vary.  We consider the limit $\beta = 8 \pi p_0/B_0^2 \gg 1$ and work
in the Boussinesq approximation, $\omega \ll k c_s$ ($c_s$ is the sound speed).
We also do not include the perturbed cosmic ray source term
(eq. [\ref{eq:crenergy}]) in our linear analysis since its form is
uncertain.

In terms of the plasma displacement,
$\bm{\xi} = \frac{i\bm{v}}{\omega}$,
the perturbed magnetic field ($\bm{B}_1 = i [\bm{k} \cdot \bm{B}_0] \bm{\xi}$) 
can be combined with equation (\ref{eq:momentum}), to give
\begin{equation}
\rho_0 \omega^2 \bm{\xi} = \frac{(\bm{k}\cdot\bm{B}_0)^2 \bm{\xi}}{4\pi}
+ i\bm{k} \Pi_1 + \rho_1 g \hat{\bm{z}},
\label{eq:acc}
\end{equation}
where 
$\Pi_1  = p_1 + p_{\rm cr 1}  + \bm{B}_0 \cdot \bm{B}_1/4\pi$
is the total-pressure perturbation. Dotting 
equation~(\ref{eq:acc}) with~$\bm{k}$ and using
the near-incompressibility condition ($\bm{k \cdot \xi} \sim \xi/H $),
we find to leading order in~$(kH)^{-1}$  that 
\begin{equation}
\Pi_1 = i\rho_1 g k_z/k^2
\label{eq:Pi1}
\end{equation}
 and
\begin{equation}
(\omega^2 - k_\parallel^2 v_A^2) \bm{\xi} = \frac{\rho_1 g}{\rho_0}
  \left(\hat{\bm{z}} - \frac{\bm{k} k_z}{k^2}\right),
\label{eq:acc2}
\end{equation}
where  $k_\parallel = \bm{k} \cdot \hat{\bm{b}}_0$, and
$v_A^2 = B_0^2/4\pi \rho_0$ is the square of the Alfv\'en speed.

In terms of the perturbation to the magnetic-field unit vector,
$\bm{\hat{b}}_1 (\equiv {\bm B}_1/B_0 - \hat{\bm b}_0 \hat{\bm b}_0 \cdot \bm{B}_1/B_0) =  i k_\parallel[ \bm{\xi}  - \hat{\bm{b}}_0(\hat{\bm{b}}_0
\cdot \bm{\xi})]$,
the perturbed heat flux can be written as
$\bm{Q}_1 = - \kappa_{\parallel,0} [
\hat{\bm{b}}_1 (\hat{\bm{b}}_0 \cdot \grad T_0)
+\hat{\bm{b}}_0  (\hat{\bm{b}}_1 \cdot \grad T_0)
+\hat{\bm{b}}_0  (\hat{\bm{b}}_0 \cdot \grad T_1)]$,
where we have dropped the term involving the perturbed
thermal conductivity since it is smaller than the
other terms by a factor of~$\sim (kH)^{-1}$.
Taking the dot product of the
heat flux with $\bm{k}$ we find that
\begin{equation}
- i \bm{k}\cdot \bm{Q}_1 = \kappa_{\parallel,0} k_\parallel^2
( 2 \xi_\parallel b_z - \xi_z) \frac{dT_0}{dz} - k_\parallel^2 \kappa_{\parallel,0} T_1,
\label{eq:Q1b}
\end{equation}
where $\xi_\parallel = \bm{\xi} \cdot \hat{\bm{b}}_0$ and
$b_z$ is the $z$ component of $ \hat{\bm{b}}_0$.
Using equations~(\ref{eq:energy}) and~(\ref{eq:Q1b}), and using
$p_1/p_0 = \rho_1/\rho_0 + T_1/T_0$, we find that
\begin{eqnarray}
\nonumber
p_1  &=& - \frac{\gamma \omega p_0 N^2 \xi_z}{(\omega+i\nu)g}
+ \left[\frac{i\nu p_0 (2\xi_\parallel b_z - \xi_z) }{\omega + i \nu}\right]
\frac{d \ln T_0}{dz} \\
&+& \frac{(\gamma \omega + i \nu)p_0 \rho_1}{(\omega + i \nu)\rho_0},
\label{eq:p1}
\end{eqnarray}
where
$\nu = {(\gamma -1) k_\parallel^2 \kappa_{\parallel,0} T_0}/{p_0}$
is the rate at which thermal conductivity smoothes
out temperature fluctuations along the magnetic field,
and 
$$
N^2= (g/\gamma) \frac{d}{dz} \ln (p_0/\rho_0^\gamma)
$$
is the square of the Brunt-V\"ais\"al\"a frequency.

In the same way that we obtained equations~(\ref{eq:Q1b})
and~(\ref{eq:p1}), for the cosmic rays we find that
\begin{eqnarray}
\nonumber
p_{\rm cr1}  &=&  
- \frac{ \gamma_{\rm cr} \omega p_{\rm cr0} M^2 \xi_z}{(\omega + i\eta)g}
+ \left[\frac{i \eta (2\xi_\parallel b_z - \xi_z)}{\omega + i\eta}\right]
 \frac{dp_{\rm cr0}}{dz} \\
&+& \frac{\gamma_{\rm cr}\omega p_{\rm cr0}\rho_1}{(\omega + i\eta)\rho_0}
\label{eq:pcr1},
\end{eqnarray}
where
$\eta = k_\parallel^2 D_\parallel$
is the rate at which diffusion smoothes out variations in
$p_{\rm cr}$ along the magnetic field, and 
$$
M^2 = (g/\gamma_{\rm cr})\frac{d}{dz} \ln(p_{\rm cr0}/\rho_0^{\gamma_{\rm cr}})
$$
is the square of the Brunt-V\"ais\"al\"a frequency associated with the cosmic ray
pressure.

Adding equations~(\ref{eq:p1}) and~(\ref{eq:pcr1}),
making use of equation~(\ref{eq:Pi1}), and noting
that the right-hand side of equation~(\ref{eq:Pi1}) is
much smaller than the individual terms on the right-hand
sides of equations~(\ref{eq:p1}) and~(\ref{eq:pcr1}),
we find that
\begin{equation}
\frac{\rho_1}{\rho_0} = \delta^{-1}
\left[ - 2i \beta^{-1} k_\parallel \xi_\parallel
+ \frac{\xi_z \overline{N}^2}{g} 
+ \frac{\omega_1^2(\xi_z - 2\xi_\parallel b_z)}{g}\right],
\label{eq:rho1}
\end{equation}
where
$$\overline{N}^2 = \gamma \omega N^2/(\omega + i \nu)
+ \gamma_{\rm cr} \alpha \omega M^2/(\omega + i \eta),$$
$\alpha = p_{\rm cr 0}/{p_0}$,
$$\omega_1^2 = g \left[
\left(\frac{i\nu}{\omega + i\nu}\right) \frac{d\ln T_0}{dz}
+ \left(\frac{i\alpha \eta}{\omega + i \eta}\right) 
\frac{d\ln p_{\rm cr0}}{dz}\right],$$
and
$$\delta = (\gamma \omega + i\nu)/(\omega + i\nu)
+\alpha \gamma_{\rm cr}\omega/(\omega + i\eta).$$

Upon substituting equation~(\ref{eq:rho1}) into
equation~(\ref{eq:acc2}), we obtain an equation for the plasma
displacement alone,
\begin{eqnarray}
\nonumber
(\omega^2 &-& k_\parallel^2 v_A^2) \bm{\xi}
- \left(\hat{\bm{z}} - \frac{\bm{k}k_z}{k^2}\right) \delta^{-1}
\left[
- \frac{2i k_\parallel \xi_\parallel g}{\beta} \right . \\
&+& \left . \xi_z \overline{N}^2 + \omega_1^2 (\xi_z - 2\xi_\parallel b_z)\right]
= 0.
\label{eq:acc3}
\end{eqnarray}
We set
$\bm{\xi} = \xi_1 \bm{\hat{e}}_1 + \xi_2 \bm{\hat{e}}_2$,
where
$\bm{\hat{e}}_1 = \bm{\hat{k}}\times \bm{\hat{z}}/{|\hat{\bm k}\times \hat{\bm z}| }$
and
$\hat{\bm e}_2 = \hat{\bm k}\times \hat{\bm e}_1$.
After taking the dot product of equation~(\ref{eq:acc3}) 
with~$\hat{\bm e}_1$ and~$\hat{\bm e}_2$, we obtain
two equations which can be written in matrix form as
\begin{equation}
\nonumber
\left(\begin{array}{cc}
\omega^2 - k_\parallel^2 v_A^2 & 0 \\
A_{21} & A_{22} 
\end{array} \right)
\left(
\begin{array}{c}
\xi_1 \\
\xi_2
\end{array}\right)
= 
\left(\begin{array}{c}
0 \\
0
\end{array}\right).
\label{eq:disp1}
\end{equation}
Setting the determinant of the matrix on the left-hand
side of equation~(\ref{eq:disp1}) equal to zero, we obtain the dispersion
relation
$(\omega^2 - k_\parallel^2 v_A^2) A_{22} = 0$,
where\footnote{
We have dropped a term on the right-hand side of
expression 
for $A_{22}$ equal to $-2i\delta^{-1}\sin^2\theta 
 k_\parallel g \beta^{-1}
(b_x k_x k_z k_\perp^{-2} - b_z)$,
which is small for $\beta \gg kH$; in the opposite limit 
$\omega = k_\parallel v_A + i \gamma$ with $\gamma/\omega 
\sim (kH)^{-1}$. Notice that $k_\parallel^2 v_A^2 : k_\parallel g \beta^{-1} 
: \overline{N}^2 :: (kH)^2\beta^{-1} : kH\beta^{-1} : 1$.}
$A_{22} = \omega^2 - k_\parallel^2 v_A^2 - \delta^{-1}\sin^2\theta 
\left( \overline{N}^2 + J\omega_1^2 \right)$,
$\theta$ is the angle between~$\bm{k}$ and $\hat{\bm z}$, 
$$
J = 1 - 2b_z^2 + {2b_x b_z k_x k_z}/{k_\perp^2},
$$ is a factor that depends on the magnetic field geometry and wavenumber, $k_\perp^2=k_x^2+k_y^2$, 
and $b_x$ is the $x$-component of~$\hat{\bm b}_0$.

While one solution to equation~(\ref{eq:disp1}), 
$\omega^2 = k_\parallel^2 v_A^2$, describes waves unaffected by buoyancy, 
$A_{22}=0$ corresponds to the modes modified by buoyancy,
\begin{equation}
\omega^2 - k_\parallel^2 v_A^2 - \delta^{-1} \sin^2\theta 
\left( \overline{N}^2 + J\omega_1^2 \right) = 0.
\label{eq:disp3}
\end{equation}
In the limit that~$ p_{\rm cr}/p \rightarrow 0$, 
equation~(\ref{eq:disp3}) reduces to equation~(13)
of \citet{qua08}.
For nonzero~$\alpha~(p_{\rm cr}/p)$,  we consider two limiting cases of 
equation~(\ref{eq:disp3}).
The first is a highly diffusive limit for cosmic rays, in which 
$\{\eta, \nu\} \gg \{\omega,N,\omega_1\} \gg k_\parallel v_A$.
In this case, 
equation~(\ref{eq:disp3}) reduces to
\begin{equation}
\omega^2 =
g J \left( \frac{d \ln T_0}{dz} + \frac{1}{p_0}\frac{d p_{\rm cr 0}}{dz}\right).
\label{eq:disp4}
\end{equation}
This is a generalization of the HBI (heat-flux driven buoyancy
instability; \citealt{qua08}) and MTI (magnetothermal instability;
\citealt{bal00}) in the presence of diffusive cosmic rays
\citep[e.g.,][]{cha06,den08}.

\begin{figure}
\centering
\epsscale{1.2}
\plotone{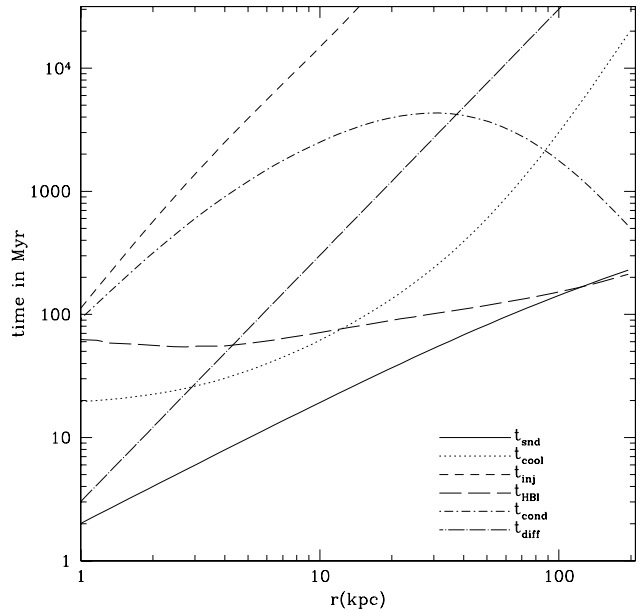}
\caption{Different timescales for the initial cluster model:
  isothermal sound crossing time (solid line, $t_{\rm snd}=r/c_s$,
  where $c_s=[p/\rho]^{1/2}$), cooling time (dotted line; we do not
  include cooling in our calculation but the cooling time is shown for
  comparison), cosmic ray energy injection time ($t_{\rm inj}=p/(\gamma-1)Q_c$;
  short-dashed line), HBI growth time ($t_{\rm HBI}=[g d\ln
  T/dr]^{-1/2}$; long-dashed line), conduction time (short dot-dashed
  line; $t_{\rm cond}=r^2 nk_B/\kappa_\parallel$), and cosmic ray
  diffusion time (long dot-dashed line; $t_{\rm
    diff}=r^2/D_\parallel$) for $D_\parallel=10^{29}$ cm$^2$s$^{-1}$.
  \label{fig:timescales}}
\end{figure}

The second limit we consider is that of adiabatic cosmic rays, with
$\nu \gg \{\omega, \overline{N}, \omega_1 \} \gg \{\eta, k_\parallel
v_A\}$.  In this case, the cosmic rays diffuse slowly compared to the
buoyancy time and behave nearly adiabatically; 
equation~(\ref{eq:disp3}) then reduces to
\begin{equation}
\omega^2 = \frac{g \sin^2 \theta}{1 + \alpha \gamma_{\rm cr}}
\left[ \frac{p_{\rm cr0}}{p_0} \frac{d}{dz}
 \ln \left( \frac{p_{\rm cr0}}{\rho_0^{\gamma_{\rm cr}}}\right)
+ J \frac{d}{dz} \ln T_0\right].
\label{eq:disp5}
\end{equation}
This implies that if the cosmic ray pressure is significant and if
cosmic ray entropy ($p_{\rm cr0}/\rho_0^{\gamma_{\rm cr}}$) is
sufficiently peaked and decreasing outward, the plasma will become
unstable to a Schwarzschild type buoyancy instability.  The entropy of
cluster plasma is stably stratified according to the Schwarzschild
criterion. The thermal plasma response at small scales is nonetheless
governed by the temperature gradient and not the entropy gradient. For
typical cluster parameters, although the global conduction timescale
may be longer than the buoyancy timescale (see
Fig. \ref{fig:timescales}), anisotropic conduction determines the
local buoyant response (i.e., $\nu \gtrsim \overline{N}$ for $kr \gg
1$) via HBI/MTI depending on the sign of temperature gradient.  The
adiabatic cosmic-ray instability (ACRI) described by equation
(\ref{eq:disp5}) is different from the MTI and HBI in that it cannot
saturate by magnetic field reorientation (the cosmic-ray driving in
eq.  [\ref{eq:disp5}] does not depend on the field configuration term
$J$ that shows up in the thermal driving). It must lead to vigorous
convection that changes the cosmic ray entropy profile to become
nearly adiabatic.

\subsection{The Cosmic Ray Diffusion Coefficient}

\label{sec:Dpar}

A cosmic ray particle streaming through a magnetized plasma is efficiently
scattered in pitch angle by magnetic fluctuations with wavelengths
comparable to its Larmor radius. An unavoidable source of
magnetic fluctuations is the self-excited streaming instability
\citep[e.g.,][]{kul69}. The effect of pitch-angle scattering due to
Alfv\'en waves is that the bulk speed of cosmic rays relative to the
thermal plasma is close to the Alfv\'en speed, i.e., $v_d-v_A=c^2/\nu
L_{\rm cr}$, where $v_d$ is the cosmic ray drift velocity relative to the thermal plasma, 
$\nu$ is
pitch-angle scattering rate, and $L_{\rm cr}$ is cosmic ray gradient
scale \citep[][]{kul05}. The cosmic rays stream along the magnetic
field direction, and down the cosmic ray pressure gradient. In addition to
streaming with Alfv\'en wave packets, cosmic rays also undergo
momentum-space diffusion, which leads to spatial diffusion along the field
lines with $D_\parallel=c^2/\nu=(v_d-v_A)L_{\rm cr}$. Thus, if cosmic
ray scattering is efficient and $v_d \approx v_A$, the diffusion
timescale over scales comparable to $L_{\rm cr}$ is much longer than
the Alfv\'en crossing time.

With the above model for cosmic ray scattering one can
self-consistently calculate the cosmic ray diffusion coefficient in
terms of the plasma parameters \citep[e.g.,][]{loe91}. In the present
calculations we do not explicitly include the effects of cosmic rays
streaming with respect to the plasma. Instead, the diffusion
coefficient in equation (\ref{eq:defGamma}) should be interpreted as
an effective diffusion coefficient, taking into account both
microscopic diffusion and streaming along turbulent magnetic field
lines. The cosmic ray ``diffusion'' time due to cosmic rays streaming
along random magnetic field lines can be crudely bounded by the
Alfv\'en crossing time ($\lesssim r/v_A$).  Together with the fact
that the true microscopic diffusion time is much longer than the
Alfv\'en crossing time (as argued above), this motivates our choice of
$D_\parallel = \alpha r v_A$ for the cosmic-ray diffusion coefficient
in most of our numerical simulations, where $\alpha$ is a factor of
order unity.  Somewhat arbitrarily, we take $\alpha = 0.4$, but our
results are insensitive to $\alpha$ so long as $\alpha \lesssim 1$.

This estimate of $D_\parallel$ ($\lesssim r v_A$) is consistent with
the measured Galactic cosmic ray diffusion coefficient for $\sim$ GeV
particles, $\sim 10^{28}$ cm$^2$s$^{-1}$ \citep{ber90}, using typical
values for magnetic field strength and cosmic ray scale height in the
Galaxy. However, at higher energies the diffusion coefficient
increases as $\epsilon^{0.5}$, where $\epsilon$ is the cosmic ray
energy (e.g., see \citealt{eng90}). For a cosmic ray energy
distribution function steeper than $\epsilon^{-2}$ (in the Milky Way
it scales as $\epsilon^{-2.7}$ from 1 to $10^5$ GeV), the cosmic ray
pressure in equation (\ref{eq:momentum}) will be dominated by the
lowest energy cosmic rays, and a diffusion coefficient $D_{\parallel}
\lesssim r v_A$ seems appropriate for the fluid description of cosmic
rays considered here. With this choice, the ratio of the buoyancy
timescale ($t^2_{\rm buoy} \sim r/g$) to the cosmic ray diffusion
timescale ($t_{\rm diff} \sim r^2/D_{\parallel}$) is $t_{\rm
  buoy}/t_{\rm diff} \lesssim \beta^{-1/2}$. In clusters $10 \lesssim
\beta \lesssim 1000$ \citep[][]{gov04} so that we expect cosmic rays
to be adiabatic on relatively large length scales and thus to be
susceptible to cosmic ray driven convection when their {\it entropy}
gradient is sufficiently large.  Our linear stability analysis differs
from that of \citet{cha06} and \citet{den08}, in that we allow for a
background heat flux.  In addition, while they obtained the dispersion
relation allowing the thermal plasma and cosmic rays to be either both
in the diffusive limit or both in the adiabatic limit, we have argued
that the more relevant case in clusters is likely that of diffusive
thermal plasma and adiabatic cosmic rays. Because of uncertainties in
$D_\parallel$, we have carried out numerical simulations for different
values of $D_\parallel$. We find that for $D_\parallel$ = 10$^{28}$
cm$^2$ s$^{-1}$, the cosmic rays are effectively adiabatic at all
scales, and even for $D_\parallel$ as high as $10^{29}$ cm$^2$
s$^{-1}$, the cosmic rays are adiabatic on large scales, $r\gtrsim 10$
kpc (see sec. 3.2.4).

\section{Numerical Simulations}

We have extended the ZEUS-MP MHD code \citep[][]{hay06,sto92a,sto92b}
to include thermal conduction along magnetic field lines
\citep[][]{sha07}, and have added cosmic rays as an additional fluid
diffusing along magnetic field lines.  We numerically solve equations
(\ref{eq:cont})-(\ref{eq:defGamma}). As mentioned earlier, we do not
include plasma cooling.  The cosmic ray energy equation, thermal
conduction in equation (\ref{eq:energy}), and the cosmic ray pressure
gradient in the equation of motion are implemented in an operator
split fashion with appropriate source and transport terms.  Both
thermal conduction and cosmic ray  diffusion are sub-cycled.

We have tested the code extensively. The anisotropic cosmic ray
diffusion equation is analogous to anisotropic thermal conduction.  We
use the method of \citet{sha07} which preserves positivity of $p_{\rm
  cr}$.  Appendix shows a 1-D shock tube test, adapted from
\citet{pfr06}, which shows that adiabatic evolution of cosmic rays is
accurate. The thermal conductivity is chosen to be the Spitzer value, \be
\label{eq:cond}
\kappa_\parallel = \frac{1.84 \times 10^{-5}}{\ln \lambda} T^{5/2}
{\rm erg s^{-1} K^{-7/2} cm^{-1}}.  \ee Based on the discussion in
\S2.2 we choose $D_\parallel = 0.4r v_A$ for most of our calculations,
where $v_A$ is the local Alfv\'en speed; to test the dependence on
$D_\parallel$, we also carry out simulations with a constant value of
$D_\parallel=10^{28},~10^{29}$ cm$^2$ s$^{-1}$.

How cosmic rays are produced and distributed in the ICM is still
poorly understood \citep[e.g.,][]{eil03}. Thus, we use a simple phenomenological source term
to drive the cosmic ray pressure in the inner parts of the cluster.
The cosmic ray energy source term in equation (\ref{eq:crenergy}) is
based on \citet{guo08}, \be
\label{eq:crsource}
Q_c = -\frac{\nu \epsilon \dot{M} c^2}{4\pi r_0^3} \left
  (\frac{r}{r_0} \right )^{-3-\nu} \left [1-e^{-(r/r_0)^2} \right].
\ee We take $r_0=20$ kpc and $\nu=1.5$, which leads to a centrally
peaked cosmic ray entropy, and take the cosmic ray energy injection
rate to be $\int Q_c 4\pi r^2 dr = 5\times 10^{42}$ erg
s$^{-1}$.\footnote{A physically more realistic model would be to
  include a feedback source term for cosmic rays where, instead of a
  fixed cosmic ray luminosity, a fixed fraction of the instantaneous
  mass accretion rate is converted into cosmic ray power; this level
  of detail is unnecessary for studying the basic physics of
  cosmic-ray driven convection but will be included in future
  calculations with radiative cooling.} Physically, the cosmic ray
energy injection rate is presumably related to the accretion rate onto
the central black hole, via $\int Q_c 4\pi r^2 dr =
\epsilon\dot{M}c^2$, where $\epsilon$ is the efficiency of cosmic ray
energy production.  Indeed, \citet{all06} show that the mechanical
luminosity of jets can be $\sim$ few \% of the inferred Bondi
luminosity.  For $\dot M = 0.1 \, M_\odot$ yr$^{-1}$, our cosmic ray
energy injection rate corresponds to $\epsilon \sim 10^{-3}$ and thus
the level of cosmic-ray power used here is observationally and
theoretically quite reasonable.

The cosmic-ray pressure fraction and its dependence on radius are not
that well-constrained observationally. For a few clusters (e.g.,
Virgo, Perseus, and Fornax) observational constraints indicate that
$p_{\rm cr}/p \lesssim 0.2$ averaged over the cluster core
\citep{pfr04,chu08}.  It is possible, however, that larger cosmic-ray
pressure fractions arise in clusters in which AGN feedback is expected
to play a particularly strong role, such as Hydra~A or Sersic 159-03
(see, e.g., \citealt{zak03,cha07}); cosmic-rays may also be
particularly important at small radii, and quite likely contribute
significantly to the pressure in and around radio bubbles (which are
associated with a deficit of X-ray emission from the thermal plasma; e.g., \citealt{bir04}).  In our simulations, we choose parameters in
equation (21) such that the cosmic ray pressure is smaller than the
plasma pressure even at late times. For a larger $Q_c$, we find that
the cosmic ray pressure builds up faster than the rate at which cosmic
ray pressure can be transported outwards via convection (or diffusion,
but the latter is slow), and the cosmic ray pressure can become larger
than the plasma pressure. In calculations with cooling, it is likely
that a larger cosmic ray injection rate could remain consistent with
$p_{\rm cr}/p \lesssim 0.25$: if the gas is allowed to cool then a
large part of the cosmic ray energy may be channeled into plasma
heating (and thus cooling) without building up a larger cosmic ray
pressure.

To assess how effectively turbulence generated by the HBI or ACRI
mixes the plasma, we solve for the advection of a passive scalar
density (e.g., a proxy for metallicity) $f$, using \be
\label{eq:passive}
\frac{df}{dt} \equiv \frac{\partial f}{\partial t} + {\bm v} \cdot
\grad f =0.  \ee Appendix \ref{app} shows the behavior of the passive
scalar density in a shock tube test.

\subsection{Simulation Parameters}

The simulations are carried out in spherical ($r$,$\theta$,$\phi$)
geometry with the inner boundary at $r_{\rm in}=1$ kpc and the outer
boundary at $r_{\rm out}=200$ kpc. Strict outflow boundary conditions
are applied to the radial velocity at the inner and outer radial
boundaries.  The plasma pressure and density are held fixed at the
outer boundary to prevent spurious oscillations; the plasma cooling
time at $r_{\rm out}$ is longer than the Hubble time. All other plasma
and cosmic ray variables are copied on ghost zones at both the inner
and outer radial boundaries. Reflective boundary conditions are
applied at the $\theta$ boundaries ($\theta=0,~\pi$).
Periodic boundary conditions are applied in the $\phi$ direction.  A
logarithmic grid is chosen in the radial direction, while the grid is
uniform in $\theta$ and $\phi$. Our fiducial run uses a $128 \times 64
\times 32$ grid, with $0 \leq \theta \leq \pi$ and $0 \leq \phi \leq
2\pi$. With these choices, $\Delta r/r=0.042$, $ \Delta \theta=0.05$,
and $\Delta \phi = 0.2$.  We also carried out a number of $128 \times
64$ 2-D (axisymmetric) simulations, and one higher resolution, $256
\times 128$, 2-D simulation for convergence studies.

We typically initialize a weak ($\beta > 10^6$ everywhere)
split-monopole magnetic field with $B \propto r^{-2}$, although in one
calculation (CRM), we use a monopole field to compare simulations with
and without net magnetic flux.  In runs with cosmic rays, the initial
cosmic ray pressure is small (0.005 times the plasma pressure at
$r_{\rm in}$) and varies as $r^{-3}$; the cosmic ray pressure builds
up in time via the source term in eq. [\ref{eq:crenergy}]).  The
initial pressure is chosen such that the plasma is in dynamical
equilibrium ($d[p+p_{\rm cr}]/dr=-\rho g$). Initial thermal
equilibrium is not imposed.

As in \citet{guo08}, we use cluster parameters relevant for Abell
2199. The gravitational potential ($\Phi$) is the sum of the dark
matter potential ($\Phi_{DM}$), \be \Phi_{DM} = -\frac{2 G M_0}{r_s}
\frac{\ln (1+r/r_s)}{r/r_s}, \ee where $M_0=3.8 \times 10^{14}
M_\odot$ is the characteristic dark matter mass, $r_s=390$ kpc is the
scale radius \citep{nav97}, and the potential due to the central cD
galaxy ($\Phi_{cD}$), \be \Phi_{cD} = -4\pi G \rho_0 r_g^2
\frac{\ln (r/r_g + \sqrt{1+(r/r_g)^2})}{r/r_g}, \ee where $r_g=2.83$
kpc, $\rho_0=5.63 \times 10^{-23}$ g cm$^{-3}$ (see \citealt{kel02}
for cD galaxy NGC 6166).

The ideal gas law $p=n k_B T$ is used with $n
\mu=n_e \mu_e=\rho/m_p$, where $\mu$ ($\mu_e$) is the mean molecular
weight per thermal particle (electron) and $n$ ($n_e$) is the total (electron) 
number density. We assume a fully ionized
plasma with hydrogen mass fraction $X=0.7$, helium mass fraction
$Y=0.28$, such that $\mu=0.62$ and $\mu_e=1.18$. Since we do not
include cooling, the metallicity appears only in the conversion of
pressure into plasma temperature used to calculate the thermal
conductivity in equation (\ref{eq:cond}). 

\begin{figure}
\centering
\epsscale{1.15}
\plottwo{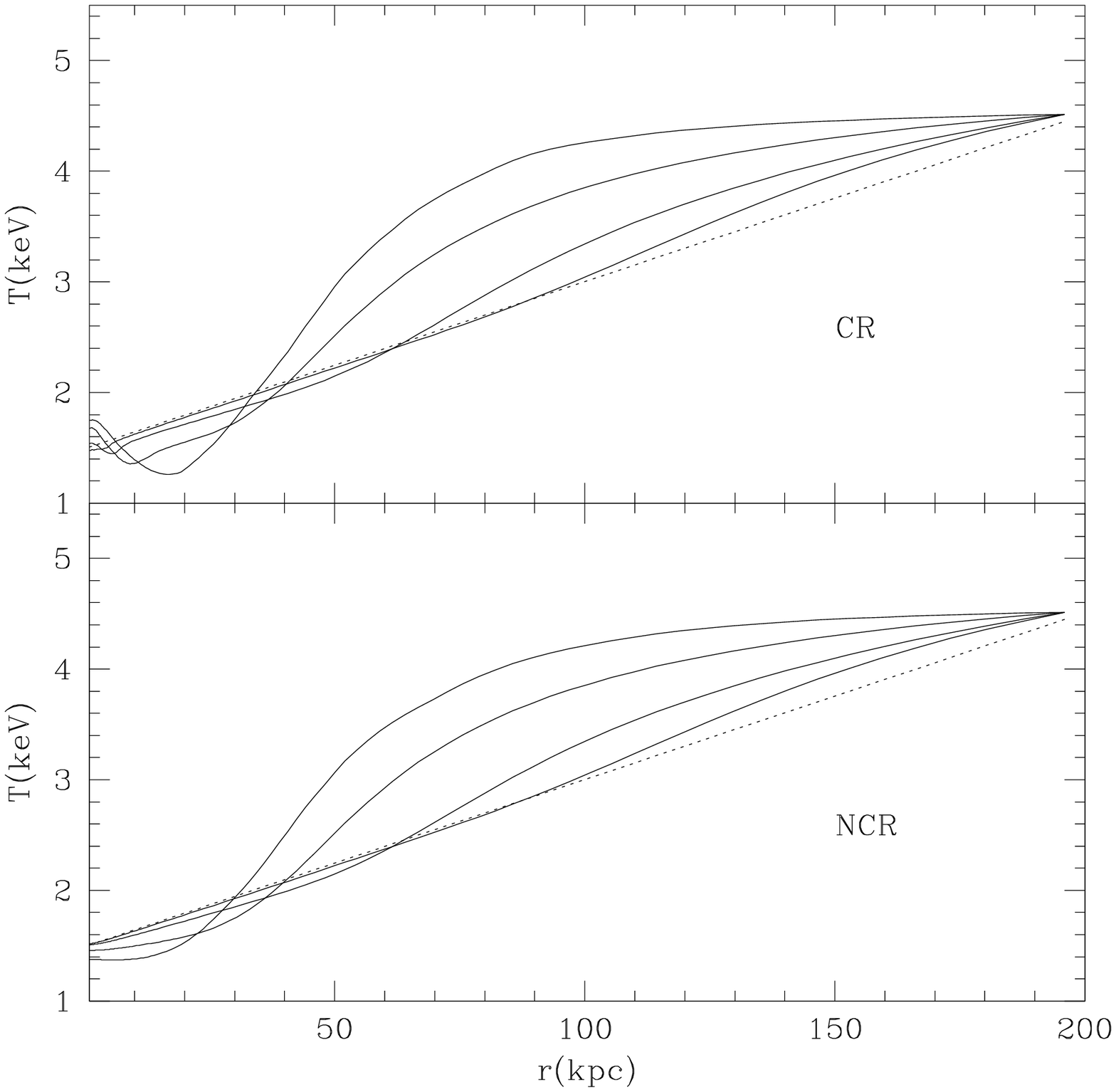}{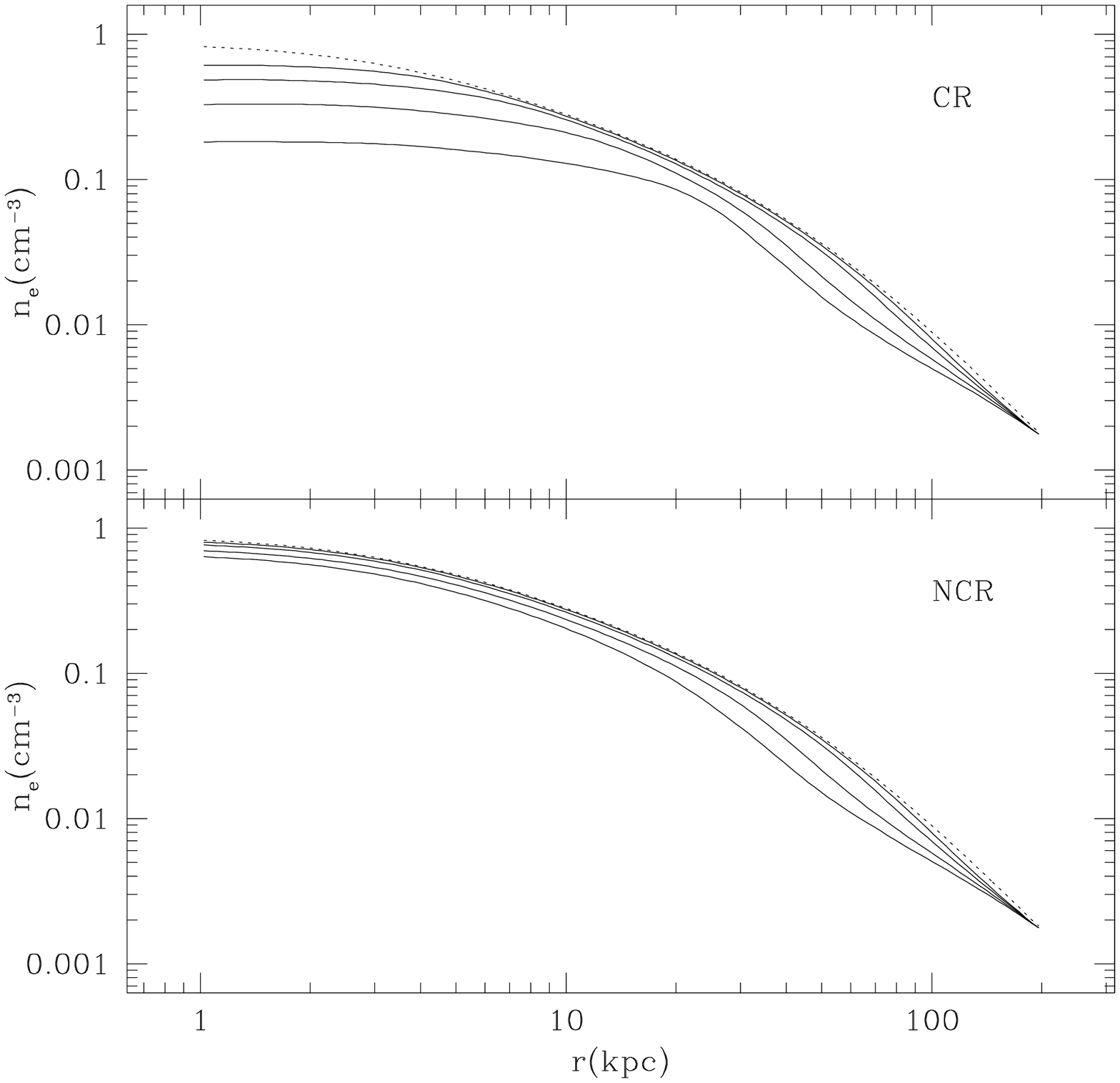}
\caption{Angle averaged plasma temperature (left) and electron number density (right) as a 
function of radius at different times for CR (upper panel) and NCR (bottom panel).
Solid lines are at 1/3, 1, 3, and 9 Gyr; dotted line is the initial profile. The temperature at $\sim 100$ kpc increases at late times while density at $\sim$ few kpc decreases with time.
\label{fig:temp}}
\end{figure}

To roughly match the observations \citep[][]{joh02}, the initial
temperature increases linearly from 1.5 keV at $r_{\rm in}$ to 4.6 keV
at $r_{\rm out}$ (see Fig. \ref{fig:temp}); the electron number
density is fixed to 0.0015 cm$^{-3}$ and the temperature is fixed to
$4.6$ keV at $r_{\rm out}$ at all times. The initial density, obtained
from imposing hydrostatic equilibrium, is quite a bit larger than the
observed density at $r_{\rm in}$.  This is because the density
obtained from hydrostatic equilibrium is extremely sensitive to the
form of the temperature profile, for which we use a simple linear fit.
However, since we do not include cooling, the large inner density does
not significantly affect our results.

\subsection{Results}
\begin{table}
\begin{center}
\caption{Parameters for different runs \label{tab:tab1}}
\begin{tabular}{ccccc}
\tableline\tableline
Label & Dim. & initial B & $D_\parallel$ & CR angle$^\ddag$\\
\tableline
CR$^\star$ &  3-D & split-M & $0.4r v_A$ & 90$^0$ \\
NCR & 3-D & split-M & 0 & 0 \\
CRM & 3-D & monopole$^\ast$ & $0.4r v_A$ &  90$^0$\\
CR2D & 2-D & split-M & $0.4 r v_A$ & 90$^0$ \\
CR2D-dbl$^\wedge$ & 2-D & split-M & $0.4 r v_A$ & 90$^0$ \\
CR28 & 2-D & split-M & $10^{28}$ & 90$^0$\\
CR29 & 2-D & split-M & $10^{29}$ & 90$^0$\\
CR30 & 3-D & split-M & $0.4r v_A$ & 30$^0$\\
CR30-ad$^\dag$ & 3-D & split-M & $0.4r v_A$ & 30$^0$\\
\tableline
\end{tabular}
\tablenotetext{$^\ddag$} {Half-angle around the polar axis over which CR source is applied.}
\tablenotetext{$\star$}{The fiducial run.}
\tablenotetext{$^\ast$}{Although monopolar, ${\bf \grad \cdot \bm{B} }=0$ everywhere, including 
the boundaries, since the origin is excluded from the computational domain.}
\tablenotetext{$^\wedge$}{Resolution for CR2D-dbl is 256 $\times$ 128, double that of CR2D.}
\tablenotetext{$^\dag$} {Plasma is adiabatic ($\kappa_\parallel=0$) for this run.}
\end{center}
\end{table}
Table \ref{tab:tab1} summarizes the properties of our
simulations. Although we list a number of calculations in Table
\ref{tab:tab1}, we focus on two 3-D simulations: CR (the fiducial run)
and NCR. Cosmic rays are not included in NCR; NCR thus serves as a
control run that allows us to isolate the effects of cosmic rays. The
aim of the rest of the simulations is to understand certain aspects of
the physics in more detail. As described earlier, for all runs except
CRM we initialize a split-monopole magnetic field. For convergence
studies, we carried out a two dimensional (axisymmetric) version of
CR, CR2D, and compared it with a run with double the resolution,
CR2D-dbl.  All runs, except CR28 and CR29, use $D_\parallel=0.4 r
v_A$; a fixed value for $D_\parallel$ is chosen for CR28 and CR29 to
test the influence of $D_\parallel$ on our results. In runs CR30 and
CR30-ad, the cosmic ray source term is only applied within 30$^0$ of
the pole to study angular diffusion of cosmic rays with (CR30) and
without (CR30-ad) thermal conduction along field lines; in the rest of
the calculations, cosmic ray injection is spherically symmetric.

Our initial profiles are in dynamical equilibrium, but not in thermal
equilibrium. The plasma will remain static if thermal conduction and
cosmic rays are not included, since the magnetic field is very
weak and the plasma is stably stratified according to the
Schwarzschild criterion. However, in the presence of thermal
conduction, the background temperature and density change in time; in
addition, the magnetic field lines are reoriented by the HBI. When the
cosmic ray source term is applied, the cosmic ray entropy gradient can
drive convection and mixing due to the ACRI. In this section we
discuss the influence of these effects on the structure of the ICM.

Figure \ref{fig:timescales} shows different timescales in the initial
state: the isothermal sound crossing time (solid line) is the shortest
timescale; the plasma cooling time (dotted line) is included for
comparison with the other timescales, although cooling is not included in
the simulations; the growth-time of the HBI (long-dashed line) varies
slowly as a function of radius; the cosmic ray injection timescale
(short-dashed line), i.e., the timescale for the cosmic-ray source
term to increase the cosmic ray pressure by an amount comparable to the
plasma pressure, increases rapidly with radius since the cosmic ray
source term is centrally concentrated; the thermal conduction
timescale (short dot-dashed line) has a maximum at intermediate radii
and is comparable to the HBI timescale at both $r_{\rm in}$ and
$r_{\rm out}$; finally, the cosmic ray diffusion timescale (long
dot-dashed line; shown for $D_\parallel=10^{29}$ cm$^2$s$^{-1}$) is
shorter than the buoyancy timescale only within $\sim 10$ kpc. Since
the diffusion time is $\propto 1/D_\parallel$, this implies that
cosmic rays are effectively adiabatic for smaller $D_\parallel$, i.e.,
in all runs except CR29 (Table \ref{tab:tab1}).

\subsubsection{The Fiducial Run (CR)}

We discuss the fiducial simulation (CR) in detail and compare it with
the simulation without cosmic rays (NCR).  Both simulations CR and NCR
show similar properties for radii $\gtrsim 30$ kpc, the radius outside
of which the cosmic ray pressure is always small.  Since there is no
cooling to balance heating by thermal conduction in the initial state,
the initial thermal properties will be modified on the conduction
timescale.

\begin{figure}
  \centering \epsscale{1.2} \plotone{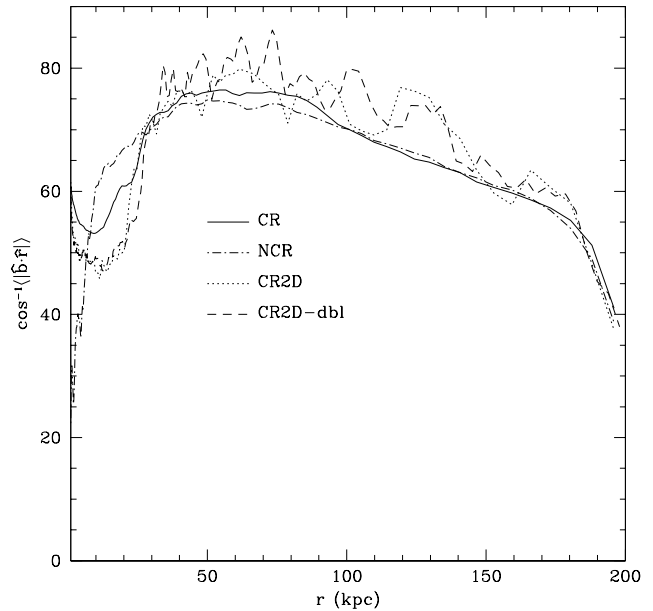}
\caption{Time (from $t_H/2$ to $3t_H/4$) and angle averaged angle between magnetic field unit vector and the radial direction 
($\cos^{-1}\lan |\hat{\bm b}\cdot \hat{\bm r}| \ran$ in degrees)  as a function of radius:  for CR (solid line),
for NCR (dot-dashed line), for CR2D (dotted line), and for CR2D-dbl (dashed line). 
\label{fig:angle}}
\end{figure}

Figure \ref{fig:temp} shows the angle averaged temperature and density
profiles as a function of radius for different times for runs CR and
NCR. Figure \ref{fig:temp} shows that for run NCR, the temperature
becomes isothermal near the inner and outer radii due to thermal
conduction along initially radial magnetic field lines. Although
Figure \ref{fig:timescales} shows that the HBI timescale is always
shorter than the conduction time, the two are of the same order at
both $r_{\rm in}$ and $r_{\rm out}$ where thermal conduction makes the
plasma isothermal before the HBI can grow significantly.  The HBI,
which is active within 30 kpc $\lesssim r \lesssim$ 100 kpc, reorients
the magnetic field lines to be primarily perpendicular to the radial
direction (see Fig. \ref{fig:angle}). This creates a thermal barrier
and a large temperature gradient at these radii. The formation of such
a thermal barrier is not forced by the boundary condition since the
temperature at $r_{\rm in}$ is floating; if the conduction timescale
were shorter than the HBI timescale at all radii, the plasma would
become isothermal (4.6 keV) at all radii. Run CR shows similar
behavior at larger radii but differs substantially at small
radii. Since the cosmic ray pressure provides a substantial fraction
of the pressure support to balance gravity within 30 kpc, the plasma
density and pressure at small radii in run CR is substantially smaller
than in NCR. The plasma temperature is similar in magnitude for CR and
NCR but shows a minimum at $\sim$ 20 kpc for CR at late times.

\begin{figure}
\centering
\epsscale{1.2}
\plotone{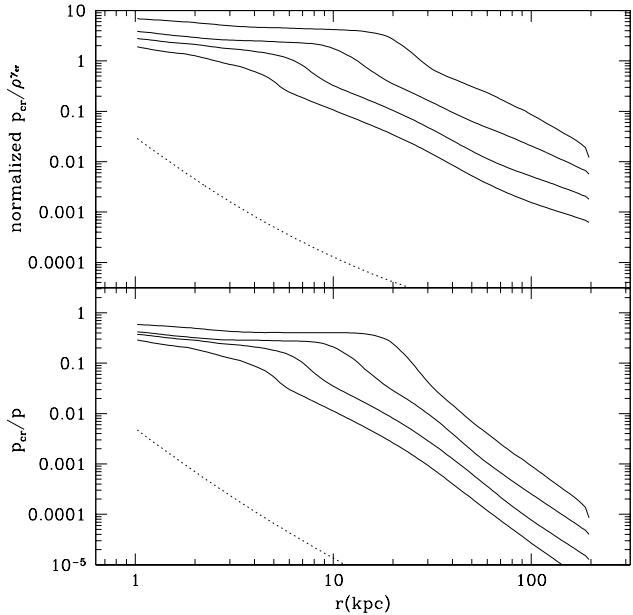}
\caption{Angle averaged cosmic ray entropy (arbitrary units; top), and the ratio of the cosmic ray pressure to the thermal plasma pressure (bottom) as a  function of radius for 1/3, 1, 3, and 9 Gyr (solid lines); dotted line is the initial profile. The ACRI flattens
  the cosmic ray entropy profile and both $p_{\rm cr}/\rho^{\gamma_{\rm cr}}$ and $p_{\rm cr}/p$ increase and move outwards in time.
  \label{fig:crent}}
\end{figure}

Figure \ref{fig:crent} shows the cosmic ray entropy profile (top) and
the ratio of cosmic ray pressure to plasma pressure ($p_{\rm cr}/p$, bottom) 
as a function of radius for different times; we define
cosmic ray entropy as $p_{\rm cr}/\rho^{\gamma_{\rm cr}}$ since this is the quantity whose
gradient determines the buoyant response of adiabatic cosmic rays
(see, e.g., eq. [\ref{eq:disp5}]).  Since the cosmic ray source term is
chosen to be a strong function of radius ($Q_c \propto r^{-2.5}$ for
$r \lesssim 20$ kpc), it drives convection due to the ACRI when the
cosmic ray pressure builds up and becomes comparable to the plasma
pressure.  At later times, cosmic ray injection does not continuously
increase the inner cosmic ray pressure with time; instead, a cosmic
ray driven convection front spreads radially outwards. Convection
drives the cosmic rays to be adiabatic ($p_{\rm cr}/\rho^{\gamma_{\rm
    cr}} \approx $ constant) in regions where the cosmic ray pressure
is not negligible compared to the plasma pressure. 

\begin{figure}
\centering
\epsscale{1.2}
\plotone{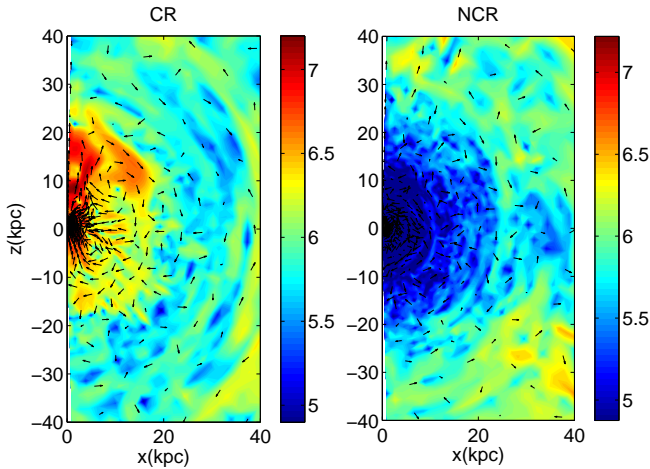}
\caption{Contour plot (in the $\phi=\pi$ plane) of $Log_{10}v$ (in cm
  s$^{-1}$) at 9 Gyr for CR (left) and NCR (right). Turbulent
  velocities $\gtrsim$ 100 km s$^{-1}$ are attained in inner 20 kpc
  for cosmic-ray driven convection (left). The turbulent velocities are significantly smaller in the absence of cosmic   rays. Arrows
  show the $r-\theta$ projection of the velocity unit vector.
  \label{fig:v}}
\end{figure}

Figure \ref{fig:v} shows 2-D contour plots of turbulent velocity
(absolute value of velocity) in the $\phi=\pi$ plane at 9 Gyr for CR
and NCR. Velocities are similar for $r \gtrsim 30$ kpc where cosmic
ray pressure is negligible and fluid motions are driven by the
HBI. The maximum turbulent velocity in the HBI-dominated regions is
$\sim 30$ km s$^{-1}$.  Run CR shows much larger turbulent velocities
($\sim$ 100 km s$^{-1}$) for $r \lesssim 30$ kpc; note that the
turbulent velocities induced by the ACRI are consistent with mixing
length theory, with $L_c \sim 4 \pi r^2 \rho v_c^3$, where $L_c$ is
the power carried by convection and $v_c$ is the resulting convective
velocity.  The large turbulent velocities in the presence of the ACRI
are sufficient to prevent catastrophic cooling according to the models
of \citet{cha07}. By contrast, the turbulent velocities are extremely
small at $r \lesssim 30$ kpc for NCR because the plasma temperature
gradient is wiped out by conduction before the HBI can drive any
turbulence.

While the turbulent velocity vectors are roughly isotropic at $r
\lesssim 30$ kpc for CR, they are aligned primarily perpendicular to
the radial direction at large radii where the HBI dominates. This is
consistent with the result that while the HBI saturates by reorienting
magnetic field lines perpendicular to gravity \citep[e.g.,][]{par08a},
the ACRI drives roughly isotropic convection irrespective of the
magnetic field geometry.

Figure \ref{fig:angle} shows the time-averaged (from $t_H/2$ to
$3t_H/4$) angle between the magnetic field vector and the radial
direction as a function of radius for runs CR and NCR; the angle is
defined with respect to the radial direction, such that it is $0^0$
for the initial split-monopole field.  The average angle is similar
for $r \gtrsim 30$ kpc, where the cosmic ray pressure is negligible,
for both simulations.  For run CR, radii $r \lesssim 30$ kpc are
convectively stirred by the ACRI and the average angle between the
magnetic field and the radial direction is $\approx 55^0$, close to
the value expected for a uniform, random magnetic field unit vector
($\cos^{-1}1/2=60^0$). For both CR and NCR at 30 kpc
$\lesssim r \lesssim 100$ kpc, the magnetic field is nearly
perpendicular to the radial direction because of the HBI.  The average
angle between the magnetic field unit vector and the radial direction
at these radii is $\approx 75^0$.  For $r \gtrsim 100$ kpc and for 
$r \lesssim 20$ kpc in run NCR, the HBI is
weak since thermal conduction wipes out the temperature gradient
before the HBI can grow significantly; as a result, the field lines
are not perpendicular to the radial direction.

\begin{figure}
\centering
\epsscale{1.2}
\plotone{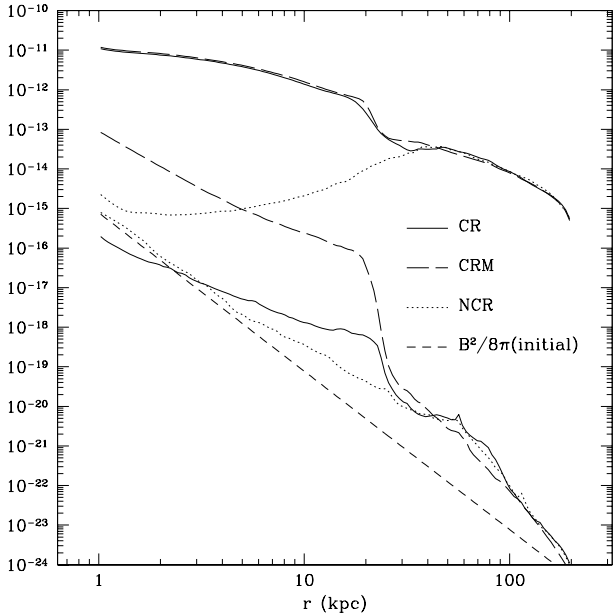}
\caption{Time (from $t_H/2$ to $3t_H/4$) and angle averaged magnetic and kinetic energy densities (in erg cm$^{-3}$) as a function of radius for CR (solid line), CRM (long dashed line), and NCR (dotted line). The kinetic energy
density is larger than the magnetic energy density in all cases. The initial magnetic energy density profile (short dashed line) is also shown for comparison. Run CRM, which includes a mean field, results in a much larger amplification of the magnetic energy as compared to CR.
\label{fig:energy}}
\end{figure}

Figure \ref{fig:energy} shows angle and time averaged (from $t_H/2$ to
$3t_H/4$) kinetic and magnetic energy profiles as a function of radius
for runs CR, CRM, and NCR.  Both the magnetic and kinetic energies are
generally amplified, although to varying degrees.  The kinetic energy
is very effectively amplified at small radii by the action of the ACRI
in run CR; the turbulent Mach number is $\sim 0.1$ (see also
Fig. \ref{fig:v}). The magnetic energy as a function of radius in CR
is quite striking in that there is no magnetic energy enhancement at
small radii: there is a slight bump in magnetic energy at 20 kpc but
the final magnetic energy is smaller than the initial magnetic energy
for $r \lesssim 3$ kpc.  In turbulent dynamos one often finds the
turbulent magnetic energy to be of the order of turbulent kinetic
energy \citep[e.g.,][]{cho00}; this is clearly not the case for the run CR. 
Magnetic field amplification can, however, be
subtle; e.g., in local shearing box simulations of the
magnetorotational instability with no net magnetic flux, magnetic
field amplification (and associated stress and turbulence) occurs only
for Prandtl numbers exceeding unity \citep[e.g.,][]{fro07}.  We do not
have explicit viscosity and resistivity and it is possible that the
effective Prandtl number in our simulation is small, so that
dissipation of the initially split monopolar field at the grid scale
dominates over magnetic field enhancement by convective turbulence. To
better understand this, we have done a simulation with cosmic rays
with a net initial magnetic flux (run CRM), but with all other
properties of the simulation the same.  Figure \ref{fig:energy} shows
that in this case, the magnetic energy density is much larger in the
inner regions as compared to CR, although the magnetic energy is still
$\sim 100$ times smaller than the kinetic energy. In this case, the
magnetic field strength in the center of the cluster is amplified to
$\sim$ 0.1-1 $\mu$G by convective motions.  The kinetic energy density
profiles are almost identical for CR and CRM, indicating that the
properties of the convection are not very different in the two cases,
although the efficiency of magnetic field amplification differs
dramatically.

At larger radii (30 kpc $\lesssim r \lesssim$ 100 kpc), the HBI causes
amplification of the magnetic and kinetic energies in both CR and
NCR. The magnetic energy is enhanced by a factor $\sim 100$ at $r
\approx 60$ kpc. This level of field amplification -- a factor of
$\sim 10$, primarily of the $\theta$ and $\phi$ components -- is what
is required to reorient initially radial magnetic fields into fields
that are perpendicular to the radial direction (as was seen in
previous HBI and MTI simulations; \citealt[][]{par08a,par07,sha08}).
The kinetic energy is also enhanced at these radii because of
turbulence driven by the HBI.

In order to study the numerical convergence of our fiducial run, we
carried out an axisymmetric 2-D run analogous to CR -- CR2D -- and an
axisymmetric run with double the resolution -- CR2D-dbl (see Table
\ref{tab:tab1}); studying convergence directly with the 3D run CR
would have been computationally prohibitive, requiring $\simeq 32$
times more cpu time.  The results from runs CR2D and CR2D-dbl are
nearly identical to each other; in particular, angle averaged plots
such as those shown in Figures \ref{fig:temp}, \ref{fig:crent},
\ref{fig:angle}, and \ref{fig:energy} are quite similar in the two
cases.  We have not included these 2-D results in every figure, but to
illustrate the basic convergence result, Figure \ref{fig:angle} shows
that the average angle between the magnetic field unit vector and the
radial direction is quite similar for CR2D and CR2D-dbl (which are
both similar to the 3-D run CR).  It is interesting to note that the
2-D runs differ from the 3-D run CR in one important respect: the
amplification of the magnetic field at small radii is significantly
larger in 2-D than in 3-D; the magnetic field energy density at small
radii in the 2-D runs is comparable to the 3-D run with a net magnetic
flux (CRM) in Figure \ref{fig:energy}.\footnote{By the anti-dynamo
  theorem, the amplification in 2D must be transient.  The dynamical
  time in clusters is so long, however, that the ``transient'' can
  last a Hubble time!} As mentioned earlier, the reason for the lack
of significant magnetic field amplification in the run CR is not
entirely clear.

\subsubsection{Diffusion of a Passive Scalar}

\begin{figure}
\centering
\epsscale{1.2}
\plotone{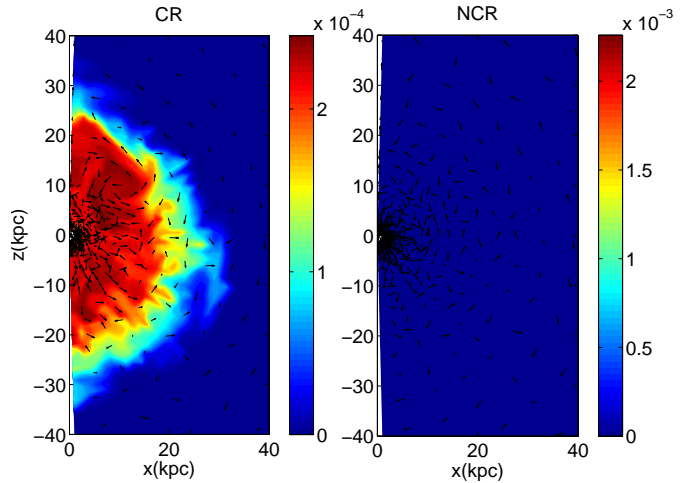}
\caption{Contour plot (at $\phi=\pi$) of the passive scalar density,
  $f$ (normalized to the initial maximum), at 9 Gyr for CR (left) and
  NCR (right). The passive scalar is initialized near the origin.  Projection of magnetic field unit vector is over-plotted
  by arrows.  While the passive scalar diffusion is negligible for
  NCR, turbulent mixing is efficient within 30 kpc for the cosmic-ray
  driven convection (left).
  \label{fig:abun}}
\end{figure}

Figure \ref{fig:abun} shows $f$, the passive scalar density, in the
inner 40 kpc for CR and NCR at 9 Gyr, as well as the projection of the
magnetic field unit vectors.  The passive scalar density is
initialized to be a large number ($f=10^{15}$) for $r<1.25$ kpc (corresponding 
to four radial zones) and is
negligible ($f=10^{-15}$) for $r>1.25$ kpc. The goal of initializing a
passive scalar is to study mixing due to turbulence.  Observations of
clusters reveal a metallicity distribution that is more spatially
extended than the light distribution of the central galaxy.  This may
indicate turbulent transport of metals in clusters
\citep[e.g.,][]{reb06,ras08}.  As expected, run CR with large
turbulent velocities at $r \lesssim 30$ kpc also results in efficient
mixing. Mixing is negligible for NCR because the inner radii ($r
\lesssim 30$ kpc) are isothermal and are thus not stirred by the HBI.
For a direct comparison with observations, one must include a time
dependent, spatially distributed source term in the passive scalar
equation which represents metal enrichment due to Type Ia supernovae
\citep[e.g.,][]{reb05}; this is beyond the scope of the present paper.
Nonetheless, our results indicate that cosmic-ray driven convection is
an efficient mechanism for mixing plasma in clusters of galaxies.

\begin{figure}
\centering
\epsscale{1.2}
\plotone{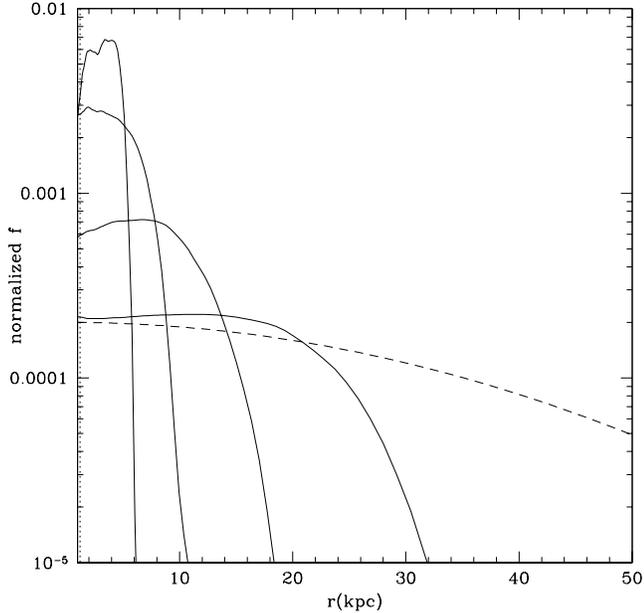}
\caption{Angle averaged passive scalar density ($Log_{10}f$;
  normalized to the initial maximum) as a function of radius for CR at
  1/3, 1, 3, 9 Gyr (solid lines); the initial profile, shown by the dotted
  line, very close to the $y$ axis. Passive scalar density decreases as
  it spreads out with time. For comparison, the dashed line shows a Gaussian fit with a diffusion coefficient $D=10^{28}$ cm$^2$s$^{-1}$ at 9 Gyr~($\sim \exp[-r^2/6Dt]$).
  \label{fig:abun1d}}
\end{figure}

Figure \ref{fig:abun1d} shows the angle averaged passive scalar
density as a function of radius for run CR at 1/3, 1, 3, 9 Gyr. Also
shown is a Gaussian fit (at 9 Gyr) with a diffusion coefficient of
$10^{28}$ cm$^2$s$^{-1}$.  The passive scalar density at 9 Gyr is
flatter than the Gaussian fit at $\lesssim$ 20 kpc, implying that
diffusion due to convection driven by cosmic rays corresponds to an
effective diffusion coefficient somewhat larger than $10^{28}$
cm$^2$s$^{-1}$. Beyond $\sim 30$ kpc, the cosmic ray pressure is
unimportant, and turbulence is driven by the HBI.  This change in the
source of the turbulence accounts for the rapid decline in the passive
scalar density at large radii in Figure \ref{fig:abun1d}. 

\begin{figure}
\centering
\epsscale{1.2}
\plotone{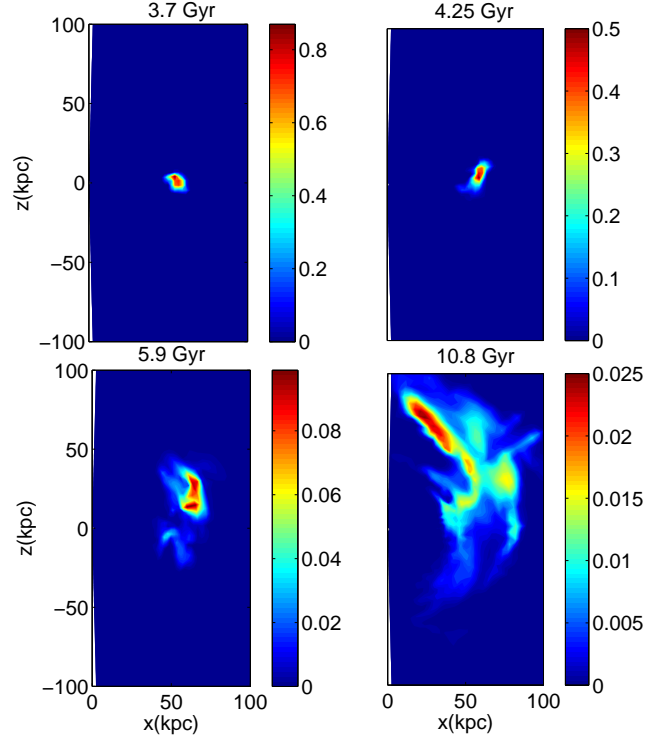}
\caption{Contour plot (at $\phi=\pi$) of the passive scalar density
  ($f$, normalized to the initial maximum) at 3.7, 4.25, 5.9, 10.8 Gyr
  for the run NCR. The passive scalar density is initialized in a small
  region (two grid points along $r, ~\theta,~\phi$) centered at 53 kpc at 3.425
  Gyr, and is negligible everywhere else. Turbulent mixing due to the HBI is faster in the $\theta$ direction compared to the radial direction.
  \label{fig:abunNCR}}
\end{figure}

To isolate just the mixing induced by turbulence driven by the HBI, we
have taken run NCR at $t_H/4(=3.425$ Gyr) and initialized a passive
scalar density peaked at $r\approx 53$ kpc, the radius where the
temperature gradient is large and the HBI is active (see
Fig. \ref{fig:temp}).  The passive scalar is initialized to be
$10^{15}$ at two grid points at $r=53$ kpc, $\theta=\pi/2$ and
$\phi=\pi$, and negligible ($10^{-15}$) elsewhere.  Figure
\ref{fig:abunNCR} shows $\phi=\pi$ snapshots of the passive scalar
density at later times.  The passive scalar diffuses more rapidly in
the $\theta$ direction.  This is because the turbulent velocities due
to the HBI are larger in the direction perpendicular to gravity than
they are in the radial direction (just as the magnetic field
components perpendicular to gravity are preferentially amplified). To
estimate the diffusion coefficient in the $r$ and $\theta$ directions,
we compare how much the passive scalar has spread in the two
directions; the full width at half maximum (FWHM) for $f$ along $r$
and $\theta$ at 9 Gyr is $\approx 10$, 40 kpc, respectively.  For
comparison, the FWHM at 9 Gyr for $f$ for run CR shown in
Fig. \ref{fig:abun1d} is $\approx 60$ kpc. Thus the diffusion
coefficient due to the HBI alone is $\sim 2$ (perpendicular to
gravity) and $\sim 50$ (parallel to gravity) times smaller than the
diffusion coefficient due to the ACRI in run CR. Although these
precise numerical values likely depend on the detailed parameters of
our simulations, the fact that the HBI primarily induces turbulence
and mixing in the plane perpendicular to gravity is a generic result.
   
\subsubsection{Heat Flux modified by the HBI and ACRI}
Many 1-D models of clusters parameterize thermal conduction by its
ratio to the Spitzer value \citep[][]{zak03,cha07,guo08}.  However,
our simulations show that, because of plasma instabilities that
operate in clusters (e.g., HBI and ACRI), a reduction of the
conductivity by a fixed factor is not applicable (see also
\citealt{par07,par08a,sha08}). At large radii in cluster cores, where
the cosmic ray pressure is negligible, the HBI can orient field lines
perpendicular to the radial direction, but at small radii where cosmic
rays can be significant, the magnetic field may be significantly more
radial. For example, for run CR the average angle of the magnetic
field relative to the radial direction for $r \lesssim 30$ kpc is
$\sim 55^0$ (see Fig. \ref{fig:angle}), corresponding (roughly) to a
reduction factor of $(\bm{\hat{b}\cdot\hat{r}})^2 \approx
1/3$. 
For 30 kpc $\lesssim r \lesssim 100$ kpc, however, the turbulence is
dominated by the HBI, and the average angle between the magnetic field
and the radial direction is $\approx 75^0$, corresponding to a
reduction factor of $\approx 0.07$.

An even more subtle result is that the HBI can change the background
temperature gradient by forming thermal barriers (see
Fig. \ref{fig:temp}), which will be absent with isotropic
conduction. For example, the temperature gradient at 30 kpc $\lesssim
r \lesssim$ 100 kpc for CR and NCR at late times is $\approx 3$ times
larger than the initial temperature gradient. Thus, the HBI not only
reduces the conductive heating by a factor of
$(\bm{\hat{b}\cdot\hat{r}})^2 \approx 0.07$, it also increases it by
making the temperature gradient larger by a factor $\approx
3$. Approximating the conductivity of a magnetized plasma by a
constant factor with respect to the Spitzer value misses all of this
interesting dynamics.  Whether this is important in real clusters with
radiative cooling and various sources of heating remains to be seen.

\subsubsection{Runs with larger $D_\parallel$ (CR28 \& CR29)}
We have also carried out 2-D simulations with larger cosmic ray
diffusion coefficients ($D_\parallel$), since the value of the cosmic
ray diffusion coefficient is poorly constrained (see \S2.2). Run CR28
uses $D_\parallel=10^{28}$ cm$^2$ s$^{-1}$, the cosmic ray diffusion
coefficient estimated for GeV cosmic rays in the Galaxy. Run CR29 uses
$D_\parallel=10^{29}$ cm$^2$ s$^{-1}$. All other parameters and
initial conditions are same as the fiducial run. Since we are
comparing these 2-D simulations with CR, which is a 3-D simulation, we
have verified that the run CR2D gives results similar to the 3-D results
presented here; in particular, the profiles for $p_{\rm
  cr}/\rho^{\gamma_{\rm cr}}$ and $p_{\rm cr}/p$ are identical to the
profiles for CR shown in Figure \ref{fig:crent}.

\begin{figure}
\centering
\epsscale{1.15}
\plottwo{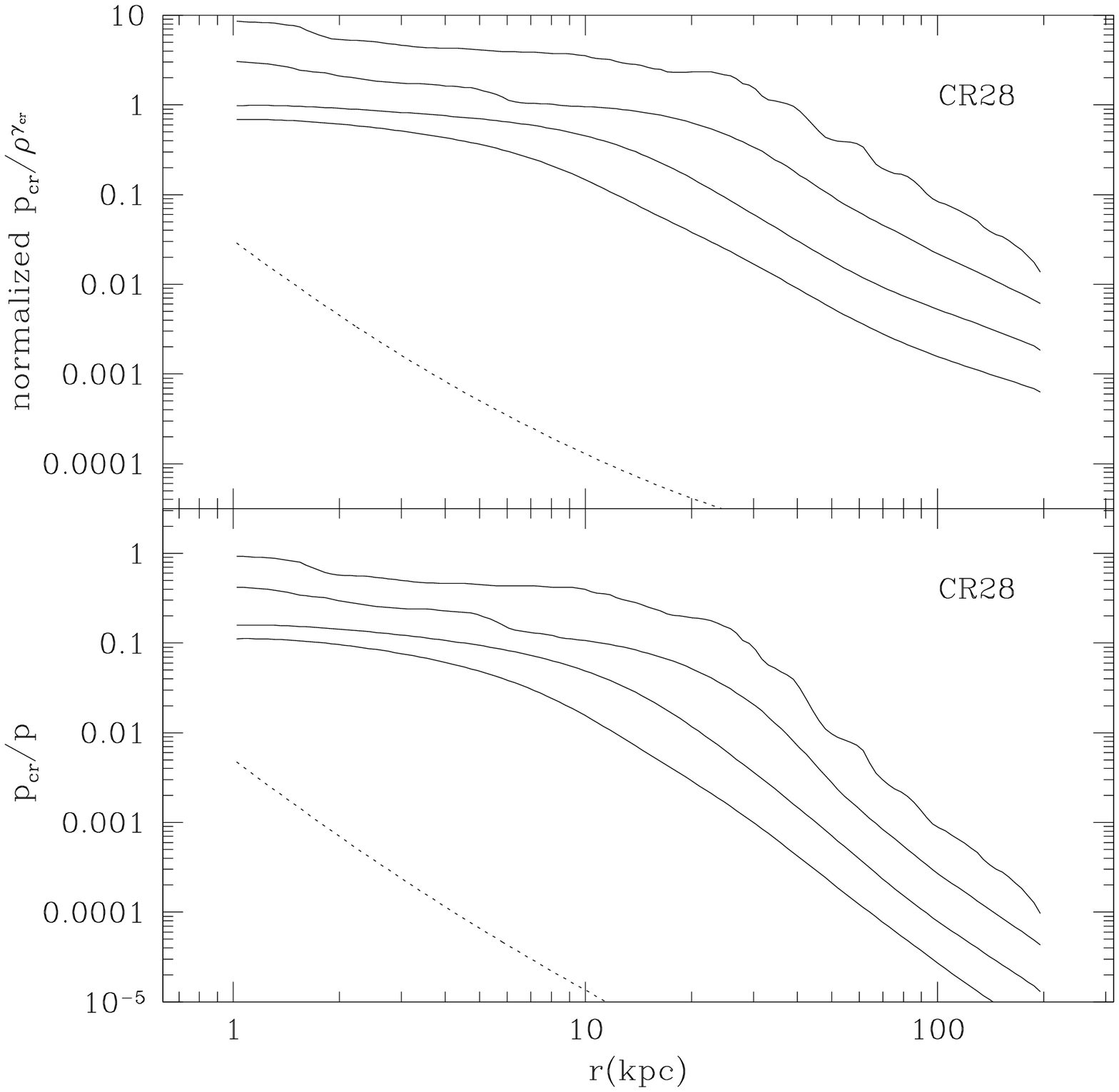}{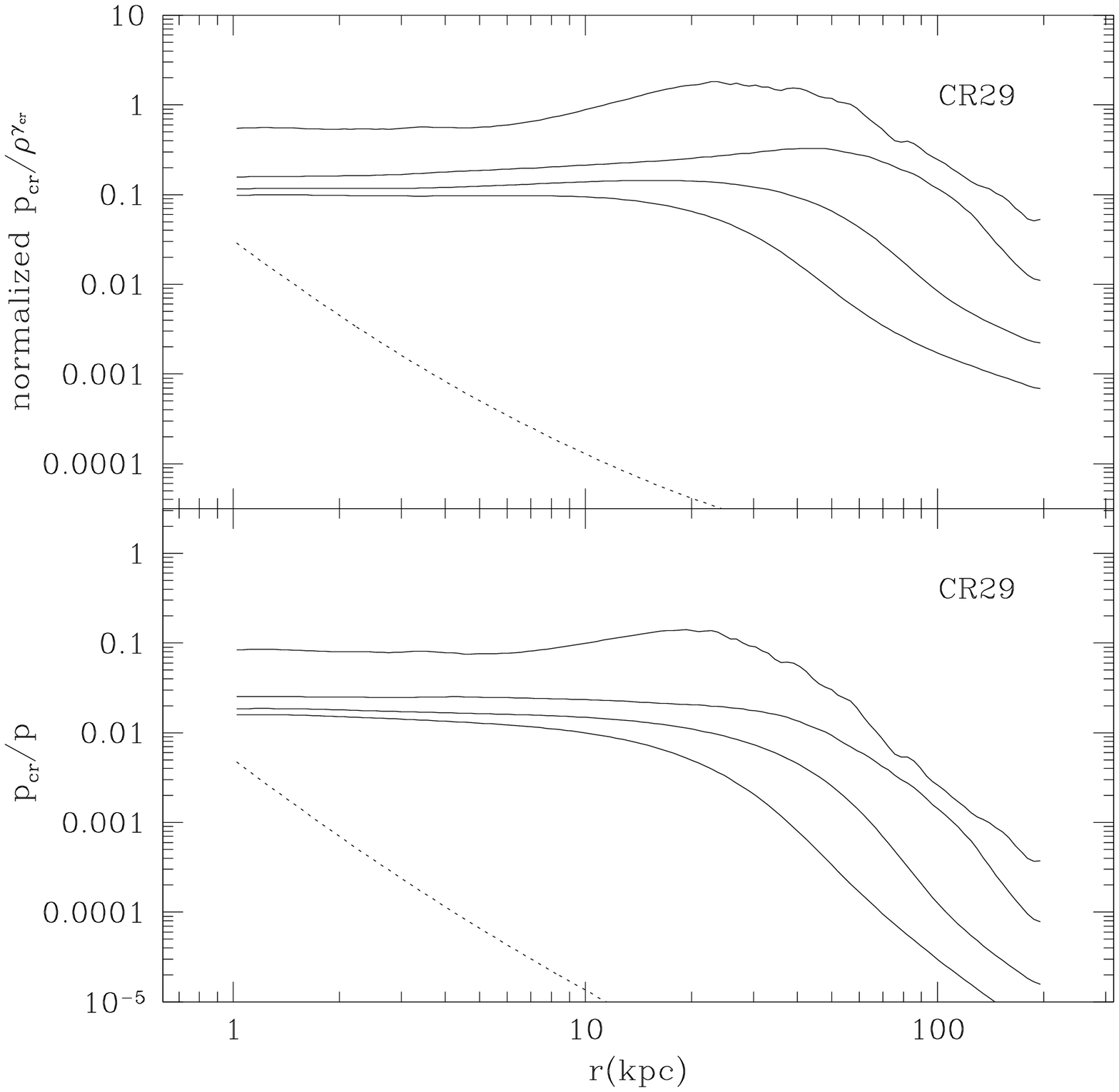}
\caption{Angle averaged cosmic ray entropy (upper panel; arbitrary
  normalization) and the ratio of cosmic ray  to plasma
  pressure (lower panel) as a function of radius for the runs CR28 (left)
  and CR29 (right), which use a fixed cosmic ray diffusion coefficients of $D_\parallel = 10^{28}$ and $10^{29}$ cm$^2$s$^{-1}$, respectively. The solid lines show profiles at 1/3, 1, 3, 9 Gyr, with the entropy and pressure ratio increasing in time;
  the initial profiles are shown with dotted lines. Profiles for CR28 look
  similar to profiles for CR in Figure \ref{fig:crent}. The profiles for CR29 are different in that $p_{\rm cr}/\rho^{\gamma_{\rm cr}}$ and $p_{\rm cr}/p$ increase
  towards a maximum at the intermediate radii; this is because the cosmic-rays are no longer adiabatic for large $D_\parallel$.
  \label{fig:largedcr}}
\end{figure}

Figure \ref{fig:largedcr} shows the angle averaged cosmic ray entropy
($p_{\rm cr}/\rho^{\gamma_{\rm cr}}$) and the ratio of the cosmic ray
pressure to the plasma pressure ($p_{\rm cr}/p$) as a function of
radius for CR28 and CR29. The profiles for CR28 and CR (see
Fig. \ref{fig:crent}) look similar; cosmic ray entropy and pressure
are slightly more radially spread out for CR28. The profiles for CR29
are quite different. The entropy and pressure ratio profiles have a
peak at $\sim 20$ kpc; this is the radius beyond which the cosmic ray
diffusion time is longer than the buoyancy timescale (see
Fig. \ref{fig:timescales}). The ratio $p_{\rm cr}/p$ is smaller in
CR29 as cosmic rays are spread out over a larger volume by
diffusion. The cosmic ray entropy increases outwards for $r \lesssim
20$ kpc since cosmic ray diffusion, and not convection driven by
cosmic rays, dominates the outward cosmic ray transport. This is
different from CR and CR28 where cosmic ray diffusion is
sub-dominant. Even in CR29, cosmic rays are effectively adiabatic for
$r \gtrsim 20$ kpc; cosmic-ray driven convection is absent at these
radii, however, because the cosmic ray pressure is not large enough to
drive the ACRI (see Fig. \ref{fig:largedcr}). At smaller radii, the
cosmic rays are nearly isobaric because of rapid diffusion, and the system is
formally unstable to the CR mediated version of the MTI
(eq. [\ref{eq:disp4}]).  However, because the field lines are nearly
radial at these radii, the growth-rate of the CRMTI is quite slow and
we do not see any indications that it develops in our simulations.

To explicitly study the possibility of the ACRI setting in at larger
radii in the cluster core, we carried out a simulation with the cosmic
ray source term (eq. [\ref{eq:crsource}]) three times larger than in
run CR29.  This larger source term increases the CR pressure at large
radii, and at late times $p_{\rm cr}/p$ is large enough to drive the
ACRI.  The turbulent velocities are $\sim 100$ km s$^{-1}$ at $r\sim
20-30$ kpc, where the cosmic rays are effectively adiabatic in spite
of the large $D_\parallel$.  Thus, even in the presence of rapid
cosmic ray diffusion, the ACRI can set in at large radii where the
cosmic rays are adiabatic, provided the cosmic ray pressure is
sufficiently large; this may naturally occur in the vicinity of
cosmic-ray filled buoyant bubbles.  More generally, our results
demonstrate that so long as $D_\parallel \lesssim 1-3 \times 10^{28}$,
cm$^2$ s$^{-1}$, cosmic rays will behave effectively adiabatically
throughout the cluster core and bulk transport by convection and other
mechanisms will dominate the diffusive transport.

\subsubsection{Runs with Cosmic ray Sources at the Poles (CR30 \& CR30-ad)}

It is very unlikely that cosmic rays in clusters are injected
spherically symmetrically.  Instead, the injection likely occurs
preferentially in the polar direction.  To study the resulting physics
in this case, we carried out 3-D simulations in which the cosmic ray
source term is applied only within $30^0$ of the pole: CR30 and
CR30-ad. Except for this difference all parameters for run CR30 are
the same as run CR. Run CR30-ad differs from CR30 in that the plasma
is adiabatic, i.e., thermal conduction is not included.  One of the
aims of these simulations is to show the dramatic differences that
result from including anisotropic thermal conduction (relative to a
more typical adiabatic simulation).  Cluster plasmas are observed to
be stable to adiabatic convection because the entropy increases
outwards \citep[e.g.,][]{pif05}. However, convection in an
anisotropically conducting plasma depends on the temperature gradient,
and not the entropy gradient, and the system is unstable independent
of the sign of the temperature gradient.  This makes it much easier to
mix a thermally conducting plasma than an adiabatic plasma.  In
clusters, this implies that turbulence produced by external means,
e.g., the ACRI, wakes due to galaxy clusters, etc., may be an
effective way of mixing the thermal plasma.

\begin{figure}
\centering
\epsscale{1.2}
\plotone{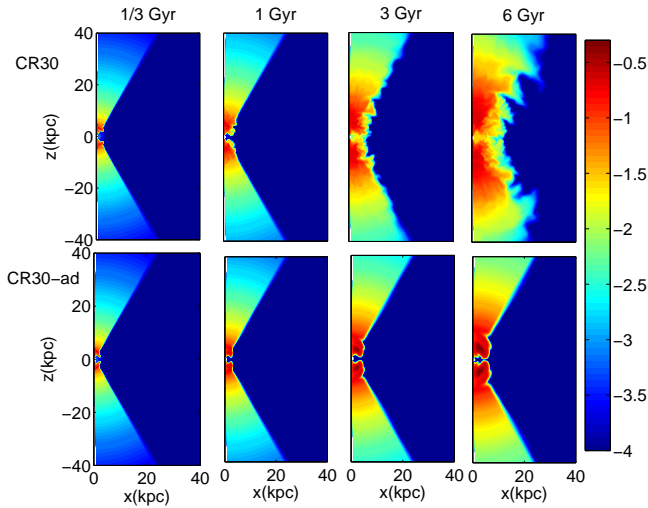}
\caption{Contour plots within 40 kpc, in the $\phi=\pi$ plane, of $Log_{10}(p_{\rm cr}/p)$ for runs CR30 (top) and CR30-ad (bottom) at $t =$ 1/3, 1, 3, and 6 Gyr (from left to right).  The ratio $p_{\rm cr}/p$ is not shown if it is smaller than $10^{-4}$.  The adiabatic plasma in CR30-ad (bottom) artificially suppresses the angular and radial mixing of the relativistic and thermal plasma that is present in the simulations with anisotropic thermal conduction (CR30; top). For movies corresponding to this figure, see: http://astro.berkeley.edu/$\sim$psharma/clustermovie.html.
\label{fig:jet}}
\end{figure}

Figure \ref{fig:jet} shows contour plots ($\phi=\pi$ snapshot) of the
ratio of cosmic ray pressure to plasma pressure ($p_{\rm cr}/p$) at
1/3, 1, 3, 6 Gyr, for CR30 and CR30-ad.  For CR30, the cosmic rays
become unstable to the ACRI in the polar region.  The resulting
turbulence is able to convectively mix the plasma, not only in the
unstable radial direction, but also in the marginally stable $\theta$
direction. Instead of cosmic rays being confined only to the
$\theta=30^0$ cone, convection effectively mixes plasma in both the
radial and angular directions.  In addition, at radii beyond 20 kpc
where the temperature gradient is appreciable (see
Fig. \ref{fig:temp}), the HBI mixes material primarily in the $\theta$
direction, as seen by the $\theta-$oriented fingers at late times in
Figure \ref{fig:jet}. For CR30, $p_{\rm cr}/p \approx 0.02$ at 3 Gyr
in the equatorial region for $r \lesssim 10-20$ kpc. Although cosmic
rays are still dominant near the pole, convection brings a
non-negligible amount cosmic rays into the equatorial region. In
comparison, there is no sign of convective overshoot in CR30-ad
because convective motions in the thermal plasma near the equator are
strongly stabilized by a large positive plasma entropy gradient. The
polar cosmic ray dominated plasma does expand somewhat in both $r$ and
$\theta$ as it becomes over-pressured.  However, the value of $p_{\rm
  cr}/p$ at 3 Gyr is $\lesssim 10^{-4}$ everywhere in the equatorial
plane for CR30-ad.

\section{Summary \& Astrophysical Implications}

The X-ray emitting plasma in clusters of galaxies is hot ($T\sim 1-10$
keV) and dilute ($n \sim 0.001-0.5$ cm$^{-3}$), so that the transport of
heat and momentum along magnetic field lines can be energetically and
dynamically important. In addition, jets launched by a central AGN
produce relativistic plasma (cosmic rays), which are observed in part
as bubbles of radio emission associated with deficits of thermal X-ray
emission (X-ray cavities; e.g., \citealt{bir04}). In this paper we have
studied the transport properties of an ICM composed of cosmic rays and
thermal plasma.  We have argued that cosmic ray diffusion is likely to
be slow because of scattering by self-generated Alfv\'en waves (\S
\ref{sec:Dpar}); as a result, the cosmic rays are adiabatic on
moderately large length scales $\gtrsim$ 1-10 kpc.
More concretely, cosmic rays are adiabatic on the scale of cluster
cores, so long as their parallel diffusion coefficient satisfies
$D_\parallel \lesssim 10^{29}$ cm$^2$s$^{-1}$.\footnote{A
  purely thermal electron-ion plasma can also show an adiabatic,
  rather than diffusive, response even in the presence of rapid electron
  thermal conduction; this occurs if electrons and protons are not
  collisionally coupled on the buoyancy timescale. We find, however,
  that even in the outer parts of clusters, the electron-proton energy
  exchange time is shorter than the buoyancy timescale and thus the
  MTI/HBI limits are appropriate.}

It is now well established that anisotropic conduction and anisotropic
cosmic ray diffusion can dramatically modify buoyancy instabilities in
low collisionality systems, producing qualitatively new instabilities
such as the MTI, HBI, and their cosmic ray counterparts
\citep[e.g.,][]{bal00,cha06,par07,qua08,par08a}.  Nonlinear studies of
these instabilities (including those in this paper) have demonstrated
that they saturate by approaching a state of marginal stability to
linear perturbations, just as in hydrodynamic convection.  However, in
a magnetized plasma, there is an additional degree of freedom that is
not present in hydrodynamic convection, namely the local direction of
the magnetic field.  The primary mechanism by which these diffusive
buoyancy instabilities saturate is by rearranging the magnetic field
lines, so that the linear growth rate becomes extremely small (see
Fig. \ref{fig:angle}). This is different from entropy gradient driven
convection in adiabatic fluids, which saturates by producing
convection that wipes out strong entropy gradients.  Even nonlinearly,
most of the energy flux in systems unstable to the MTI and HBI is
transported by thermal conduction, rather than convection.  Moreover,
the saturation of these instabilities is quasilinear in the sense that
the saturated magnetic energy is proportional to the initial magnetic
field energy \citep{sha08}.

In this paper we have shown analytically and through numerical
simulations that when cosmic rays have appreciable pressure, $p_{\rm
  cr}/p\gtrsim 0.25$, and an outwardly decreasing entropy ($p_{\rm
  cr}/\rho^{\gamma_{\rm cr}}$), they can drive strong convection and
mixing in clusters of galaxies.  This adiabatic cosmic ray instability
(ACRI) in the central regions of clusters of galaxies is a cosmic ray
analogue of hydrodynamic convection familiar in the context of stars
and planets.  In particular, the nonlinear saturation of adiabatic
cosmic ray convection is similar to that of hydrodynamic convection,
and thus quite different from the saturation of the MTI and HBI.  Our
simulations of cluster cores also provide insight into the global
saturation of the HBI at radii in clusters where the cosmic ray
pressure is negligible ($\sim 30-100$ kpc in our models), and thus
where the only convective instability is that driven by the background
conductive heat flux.  More specifically, the primary results of this
paper include:

\begin{itemize}
\item If the cosmic ray entropy decreases outwards and if $p_{\rm
    cr}/p \gtrsim 0.25$, convection driven by the ACRI sets in.  In
  the saturated state, the cosmic ray entropy profile becomes nearly
  constant in the region with significant cosmic ray pressure
  (Fig. \ref{fig:crent}).  The resulting turbulent velocities are
  consistent with mixing length theory, with $v_c \sim 100
  (L_c/10^{43} \, {\rm erg \, s^{-1}})^{1/3} \, (n/0.1 \, {\rm
    cm^{-3}})^{-1/3} (r_0/20 {\rm kpc})^{-2/3}$ km s$^{-1}$ where $L_c$ is 
   the total power
  supplied to cosmic rays and $r_0$ is the pressure height scale of 
  cosmic rays. The ACRI generates turbulent motions more
  effectively in cluster cores than the HBI alone.

\item The ACRI drives roughly isotropic convection with the average
  angle between the field lines and the radial direction $\sim 55^0$;
  by contrast, the HBI generates magnetic field lines that are
  primarily in the $\theta$ and $\phi$ directions
  (Fig. \ref{fig:angle}), shutting off the radial conduction of heat.
  The effective radial conductivity of a cluster plasma thus depends
  sensitively on which of these instabilities operates at a given
  location, and may not be adequately approximated as a fixed fraction of the
  Spitzer value throughout the cluster.

\item We have quantified the mixing of a passive scalar by the ACRI
  and HBI: the ACRI produces roughly isotropic mixing with a turbulent
  diffusion coefficient $D \gtrsim 10^{28}$ cm$^2$s$^{-1}$
  (Fig. \ref{fig:abun}); mixing length theory predicts that $D \propto
  v_c \propto L_c^{1/3}$.  At larger radii, only the HBI operates and
  the mixing is primarily in the $\theta$ and $\phi$ directions,
  rather than in the radial direction (Fig. \ref{fig:abunNCR}).  Both the
  ACRI and the HBI may contribute to mixing metals in clusters 
  by redistributing, in both radius and angle, metals
  produced by Type 1a supernovae.  Some observations of metallicity
  gradients in clusters have inferred mixing at levels comparable to
  those found here \citep[e.g.,][]{reb06}.

\item It is considerably easier to mix thermal plasma in the presence
  of anisotropic thermal conduction, since the plasma is formally
  always buoyantly unstable and thus already prone to mixing!  By
  contrast, treating the plasma as adiabatic (i.e., ignoring thermal
  conduction) results in an {\it artificially stabilizing} entropy
  gradient in cluster plasmas.\footnote{Even including isotropic
    thermal conduction reduces the stabilizing effect of the entropy
    gradient; it does not, however, capture the MTI/HBI, which are
    driven by anisotropic thermal conduction along magnetic field
    lines.}  As a concrete example of these effects, we have
  demonstrated that cosmic rays initially injected into the polar
  regions can be partially mixed to the equator by convective
  overshooting in the ACRI and HBI unstable regions
  (Fig. \ref{fig:jet}); this effect is largely absent in simulations
  that treat the plasma as adiabatic.  If wave heating due to cosmic
  ray streaming or heating due to Coulomb interactions is important in
  clusters (e.g., \citealt{guo08}), a mechanism similar to that
  described here may be crucial in redistributing cosmic rays
  throughout the cluster volume.  More generally, to study the mixing
  produced by external sources of turbulence such as galactic wakes or
  cosmic-ray filled bubbles, we suspect that anisotropic thermal
  conduction must be accounted for, so that the buoyant response of
  the thermal plasma is correctly represented.

\end{itemize}

Having summarized our primary results, we now describe several caveats
and directions for future research.  First, to generate the ACRI, we
have injected cosmic rays using a subsonic source term at small
radii. In reality a significant fraction of the cosmic rays produced
by an AGN are expected to be produced in a supersonic jet shocking
against the ICM. The spatial distribution of cosmic rays produced by
jets is poorly understood. The intuition drawn from our simulations
should apply as long the source of cosmic rays ultimately produces a
centrally concentrated bubble of relativistic plasma that expands
subsonically.

Our calculations intentionally do {\em not} include plasma
cooling. Instead of trying to solve the cooling flow problem, our goal
has been to study the basic physics of buoyancy instabilities in the
combined relativistic + thermal plasma, implicitly assuming that some
heating process is preventing catastrophic cooling of the plasma.  Our
calculations also do not include anisotropic ion viscosity which is
$\approx 40$ times smaller than electron thermal conductivity. 
Finally, we do not treat the effects of cosmic ray streaming with
respect to the thermal plasma from first principles, although our
choice of the cosmic ray diffusion coefficient qualitatively accounts
for limits on cosmic ray streaming produced by self-excited Alfv\'en
waves (see \S \ref{sec:Dpar}).  In future work, we intend to include
all of the above effects, which will provide a more quantitative model
of plasma in cluster cores.

Finally, we note that in a full cosmological context, galaxy clusters
will contain a large number of galaxies and other dark matter
substructure.  The motion of such bound objects through the ICM will
reorient the magnetic field and generate downstream turbulence.  The
interplay between this turbulence and that generated by the
instabilities studied in this paper is worth investigating in detail
in future work.  This interaction may create a magnetic dynamo in the
ICM that is more effective than that produced by the HBI alone:
galaxies moving through the ICM will comb out the magnetic field lines
in the radial direction, while the HBI will amplify the field and
generate a strong perpendicular magnetic field component from the seed
radial field created by galactic wakes.

\acknowledgements Support for this work was provided by NASA through
Chandra Postdoctoral Fellowship grant numbers PF8-90054 and PF7-80049
awarded by the Chandra X-ray Center, which is operated by the
Smithsonian Astrophysical Observatory for NASA under contract
NAS8-03060.  E. Q. was supported in part by the David and Lucile
Packard Foundation, NSF-DOE Grant PHY-0812811, and NSF
ATM-0752503. B. C. was supported in part by NASA grant No. NNG 05GH39G
and NSF grant No. AST 05-49577. We thank the Laboratory for
Computational Astrophysics, University of California, San Diego, for
developing ZEUS-MP and providing it to the community. This research
was supported in part by the National Science Foundation through
TeraGrid resources provided by NCSA and Purdue University. The
simulations reported in the paper were carried out on the Abe cluster
at NCSA and the Steele cluster at Purdue University.

\appendix 
\section{Numerical Tests}
\label{app}
\begin{figure}
\centering
\epsscale{0.8}
\plotone{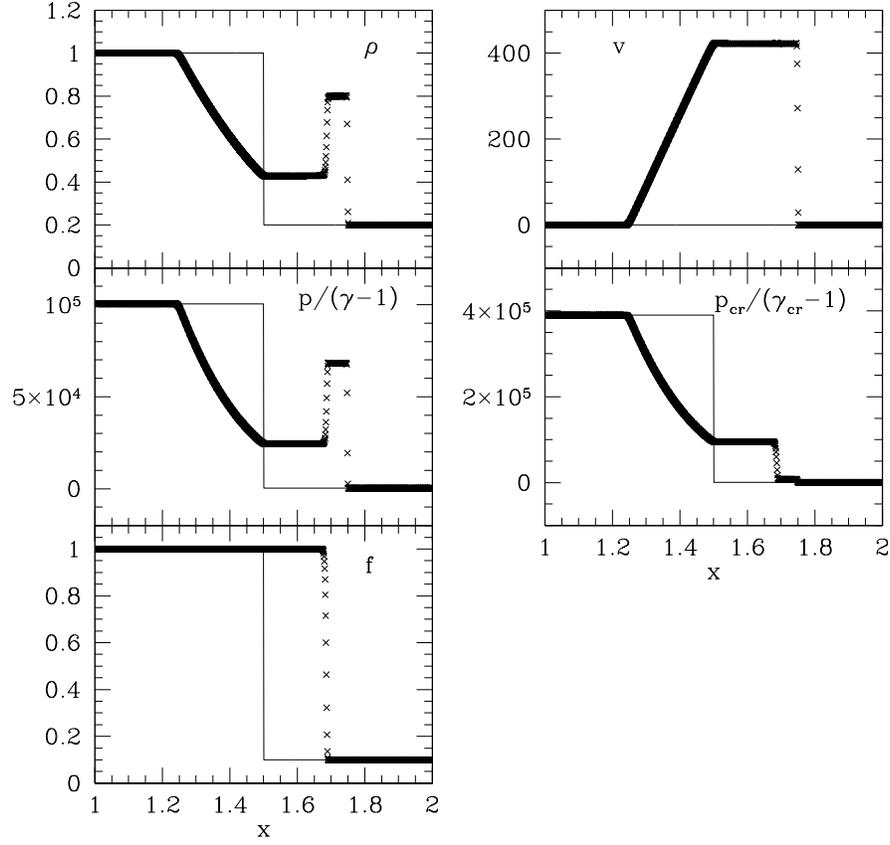}
\caption{Profiles for different fluid variables for the shock tube test: initial profiles (solid line) and profiles at 
$t=4.4\times 10^{-4}$ (points). While the shock is resolved by 4 points, contact discontinuity
requires more grid points to be resolved (this is a feature of all methods that do not solve 
the full Riemann problem).
\label{fig:app}}
\end{figure}
\subsection{Cosmic ray shock tube}
We have tested the adiabatic implementation of cosmic rays with a 1-D shock tube problem 
discussed in 
\citet{pfr06} and in \citet{ras08}. Like \citet{ras08}, we also use 1024 grid points.
Thermal conduction and cosmic ray diffusion are absent for
this test problem. 
The left ($1<x<1.5$) and right ($1.5 \leq x < 2$) states are given by 
($\rho_L$, $v_L$, $p_L$, $p_{crL}$) = (1, 0, $6.7\times10^4$, $1.3 \times 10^5$) and 
($\rho_R$, $v_R$, $p_R$, $p_{crR}$) = (0.2, 0, $2.4\times10^2$, $2.4 \times 10^2$), respectively. Figure \ref{fig:app} shows profiles at $t=0$ (solid line) and at $t=4.4\times 10^{-4}$ (shorter than the crossing time; points). The profiles match very well with the analytic result and with Fig. 3 of \citet{ras08}. In addition we also test advection of a passive scalar density governed by equation (\ref{eq:passive}). The passive scalar density shows a discontinuity at the location of the contact discontinuity; volume integrated $f$ is not conserved but volume integrated $\rho f$ is.

\end{document}